\documentclass[11pt, notitlepage]{article}
\usepackage{latexsym}
\usepackage[dvips]{graphics,color}
\def\be{\begin{displaymath}}
\def\ee{\end{displaymath}}
\def\bne{\begin{equation}}
\def\ene{\end{equation}}
\def\bee{\begin{eqnarray*}}
\def\eee{\end{eqnarray*}}
\def\bnee{\begin{eqnarray}}
\def\enee{\end{eqnarray}}

\font\openface=msbm10 at10pt
\def\past{\mathrm{past}}

\def\nc2{{N \choose 2}}
\def\Nc2{{N \choose 2}} % too wasteful?

\def\half{\frac{1}{2}}
\def\quart{\frac{1}{4}}
\def\thquart{\frac{3}{4}}
\def\E#1{<\!#1\!>}
\def\interval{\mathrm{int}}

\def\lto{\mathop
        {\hbox{${\lower3.8pt\hbox{$<$}}\atop{\raise0.2pt\hbox{$\sim$}}$}}}
\def\gto{\mathop
        {\hbox{${\lower3.8pt\hbox{$>$}}\atop{\raise0.2pt\hbox{$\sim$}}$}}}

\def\eprint #1 {$\langle$e-print arXive: #1$\rangle$}

\font\openface=msbm10 at11pt

\def\NaturalNumbers{\hbox{\openface N}} 
\def\Minkowski{\hbox{\openface M}}

\def\=>{\Rightarrow}
\def\==>{\Longrightarrow}

 \def\dal{\displaystyle{{\hbox to 0pt{$\sqcup$\hss}}\sqcap}}

\def\lto{\mathop
        {\hbox{${\lower3.8pt\hbox{$<$}}\atop{\raise0.2pt\hbox{$\sim$}}$}}}
\def\gto{\mathop
        {\hbox{${\lower3.8pt\hbox{$>$}}\atop{\raise0.2pt\hbox{$\sim$}}$}}}
\def\half{{1\over 2}}

\def\to{\rightarrow}		% symbol for `approaches' or `maps to'

\def\bar{\overline}		% define bar to always be wide bar
\def\interior #1 {  \buildrel\circ\over  #1}     % seems to work
\def\twoch{\,\,\begin{picture}(0,1) % ``2-chain''
\thicklines
\multiput(0,0)(0,1){2}{\circle*{.6}}
\put(0,0){\line(0,1){1}}
\end{picture}\,\,}

\def\twoach{\,\,\begin{picture}(1,0) % ``2-antichain''
\thicklines
\multiput(0,0)(1,0){2}{\circle*{.6}}
\end{picture}\,\,}

\def\threech{\,\,\begin{picture}(0,2) % threech ``3-chain''
\thicklines
\multiput(0,0)(0,1){3}{\circle*{.6}}
\put(0,0){\line(0,1){2}}
\end{picture}\,\,}

\def\wedge{\,\,\begin{picture}(2,2) % wedge ``Lambda''
\thicklines
\put(1,2){\circle*{.6}}
\multiput(0,0)(2,0){2}{\circle*{.6}}
\put(0,0){\line(1,2){1}}
\put(2,0){\line(-1,2){1}}
\end{picture}\,\,}

\def\Lcauset{\,\,\begin{picture}(2,2) % Lcauset ``L''
\thicklines
\multiput(0,0)(1.5,0){2}{\circle*{.6}}
\put(0,2){\circle*{.6}}
\put(0,0){\line(0,1){2}}
\end{picture}\,\,}

\def\V{\,\,\begin{picture}(2,2) % V ``V''
\thicklines
\put(1,0){\circle*{.6}}
\multiput(0,2)(2,0){2}{\circle*{.6}}
\put(1,0){\line(-1,2){1}}
\put(1,0){\line(1,2){1}}
\end{picture}\,\,}

\def\threeach{\,\begin{picture}(2,1) % threeach ``3-antichain''
\thicklines
\multiput(0,.5)(1,0){3}{\circle*{.6}}
\end{picture}\,\,}

\def\fourch{\,\,\begin{picture}(0,3) % fourch ``4-chain''
\thicklines
\multiput(0,0)(0,1){4}{\circle*{.6}}
\put(0,0){\line(0,1){3}}
\end{picture}\,\,}

\def\fourach{\,\,\begin{picture}(3,1) % fourach ``4-antichain''
\thicklines
\multiput(0,.5)(1,0){4}{\circle*{.6}}
\end{picture}\,\,}

\def\Y{\,\,\begin{picture}(2,2) % Y
\thicklines
\multiput(0,0)(0,1){3}{\circle*{.6}}
\put(2,2){\circle*{.6}}
\put(0,0){\line(0,1){2}}
\put(0,1){\line(2,1){2}}
\end{picture}\,\,}

\def\iY{\,\,\begin{picture}(2,2) % iY
\thicklines
\multiput(0,0)(0,1){3}{\circle*{.6}}
\put(2,0){\circle*{.6}}
\put(0,0){\line(0,1){2}}
\put(0,1){\line(2,-1){2}}
\end{picture}\,\,}

\def\threecho{\,\,\begin{picture}(1,2) % threecho ``31''
\thicklines
\multiput(0,0)(0,1){3}{\circle*{.6}}
\put(1,0){\circle*{.6}}
\put(0,0){\line(0,1){2}}
\end{picture}\,\,}

\def\Lo{\,\,\begin{picture}(2,1) % Lo ``L1''
\thicklines
\multiput(0,0)(1,0){3}{\circle*{.6}}
\put(0,1){\circle*{.6}}
\put(0,0){\line(0,1){1}}
\end{picture}\,\,}

\def\new{\,\,\begin{picture}(2,2) % new ``nu''
\thicklines
\multiput(0,0)(0,1){3}{\circle*{.6}}
\put(2,1){\circle*{.6}}
\put(0,0){\line(0,1){2}}
\put(0,0){\line(2,1){2}}
\end{picture}\,\,}

\def\inu{\,\,\begin{picture}(2,2) % inu
\thicklines
\multiput(0,0)(0,1){3}{\circle*{.6}}
\put(2,1){\circle*{.6}}
\put(0,0){\line(0,1){2}}
\put(2,1){\line(-2,1){2}}
\end{picture}\,\,}

\def\diamond{\,\,\begin{picture}(2,2) % diamond
\thicklines
\multiput(0,0)(0,1){3}{\circle*{.6}}
\put(2,1){\circle*{.6}}
\put(0,0){\line(0,1){2}}
\put(0,0){\line(2,1){2}}
\put(2,1){\line(-2,1){2}}
\end{picture}\,\,}

\def\flower{\,\,\begin{picture}(2,2) % flower
\thicklines
\multiput(0,2)(1,0){3}{\circle*{.6}}
\put(1,0){\circle*{.6}}
\put(1,0){\line(-1,2){1}}
\put(1,0){\line(0,1){2}}
\put(1,0){\line(1,2){1}}
\end{picture}\,\,}

\def\iflower{\,\,\begin{picture}(2,2) % iflower
\thicklines
\multiput(0,0)(1,0){3}{\circle*{.6}}
\put(1,2){\circle*{.6}}
\put(0,0){\line(1,2){1}}
\put(1,0){\line(0,1){2}}
\put(2,0){\line(-1,2){1}}
\end{picture}\,\,}

\def\wedgeo{\,\begin{picture}(3,2) % wedgeo ``Lambda + 1''
\thicklines
\put(1,2){\circle*{.6}}
\put(3,0){\circle*{.6}}
\multiput(0,0)(2,0){2}{\circle*{.6}}
\put(0,0){\line(1,2){1}}
\put(2,0){\line(-1,2){1}}
\end{picture}\,\,\,}

\def\Vo{\,\begin{picture}(3,2) % Vo ``V + 1''
\thicklines
\multiput(0,2)(2,0){2}{\circle*{.6}}
\multiput(1,0)(2,0){2}{\circle*{.6}}
\put(1,0){\line(-1,2){1}}
\put(1,0){\line(1,2){1}}
\end{picture}\,\,}

\def\pie{\,\,\begin{picture}(1,1) % pie ``Pi''
\thicklines
\multiput(0,0)(0,1){2}{\circle*{.6}}
\multiput(1,0)(0,1){2}{\circle*{.6}}
\put(0,0){\line(0,1){1}}
\put(1,0){\line(0,1){1}}
\end{picture}\,\,}

\def\N{\,\,\begin{picture}(1,2) % N
\thicklines
\multiput(0,0)(0,2){2}{\circle*{.6}}
\multiput(1,0)(0,2){2}{\circle*{.6}}
\put(0,0){\line(0,1){2}}
\put(1,0){\line(0,1){2}}
\put(0,0){\line(1,2){1}}
\end{picture}\,\,}

\def\bowtie{\,\,\begin{picture}(2,2) % bowtie
\thicklines
\multiput(0,0)(0,2){2}{\circle*{.6}}
\multiput(2,0)(0,2){2}{\circle*{.6}}
\put(0,0){\line(0,1){2}}
\put(2,0){\line(0,1){2}}
\put(0,0){\line(1,1){2}}
\put(2,0){\line(-1,1){2}}
\end{picture}\,\,}

\def\interval{{\rm int}}
\def\Minkowski{\hbox{\openface M}}
\def\E#1{<\!#1\!>}

\begin{document}
\title{Evidence for a continuum limit in causal set dynamics}
\author{\em D. P. Rideout \thanks{rideout@physics.syr.edu} 
	\, and
	R. D. Sorkin \thanks{sorkin@physics.syr.edu} \\
	 \em Department of Physics, Syracuse University \\ 
	\em Syracuse, NY, 13244-1130, U.S.A.}
\maketitle

%: abstract

\begin{abstract}
We find evidence for a continuum limit of a particular causal set
dynamics which depends on only a single ``coupling constant'' $p$ and
is easy to simulate on a computer.  
% Such simulations indicate that, as the number $N$ of causal set elements
% goes to infinity, one can adjust $p$ so as to hold constant the coarse
% grained physics at any fixed scale.
The model in question is a stochastic process that can also be
interpreted as 1-dimensional directed percolation, or in terms of
random graphs.
\end{abstract}

%: introduction

\section{Introduction}

%:: we defined  a dynamics earlier

In an earlier paper \cite{class_dyn} we investigated a type of causal
set dynamics that can be described as a (classically) stochastic
process of growth or ``accretion''.  In a language natural to that
dynamics, the passage of time consists in the continual birth of new
elements of the causal set and the history of a sequence of such
births can be represented as an upward path through a poset of all
finite causal sets.  We called such a stochastic process a {\it
sequential growth dynamics} because the elements arise singly, rather
than in pairs or larger multiplets.

A sequential description of this sort is advantageous in representing
the future as developing out of the past, but on the other hand it
could seem to rely on an external parameter time (the ``time'' in
which the growth occurs), thereby violating the principle that
physical time is encoded in the 
intrinsic
order-relation of the causal
set and nothing else.  If physically real, such a parameter time would
yield a distinguished labeling of the elements and thereby a notion of
``absolute simultaneity'', in contradiction to the lessons of both
special and general relativity.  To avoid such a consequence, we
postulated a principle of {\it discrete general covariance}, according
to which no probability of the theory can depend on --- and no physically
meaningful question can refer to --- the imputed order of births, except
insofar as that order reflects the intrinsic precedence relation of the
causal set itself.

To discrete general covariance, we added two other principles that we
called {\it Bell causality} and {\it internal temporality}.  The first
is a discrete analog of the condition that no influence can propagate
faster than light, and the second simply requires that no element be
born to the past of any existing element.\footnote%
{This last condition guarantees that the ``parameter time'' of our
 stochastic process {\it is compatible with} physical temporality, as
 recorded in the order relation $\prec$ that gives the causal set its
 structure.  In a broader sense, general covariance itself is also an
 aspect of internal temporality, since it guarantees that the
 parameter time {\it adds nothing to} the relation $\prec$.}
These principles led us almost uniquely to a family of dynamical laws
(stochastic processes) parameterized by a countable sequence of
coupling constants 
$q_n$.  
In addition to this generic family,
there are some exceptional families of solutions, 
but we conjecture that they are all singular limits of the generic family.  
We have checked in particular that ``originary percolation'' 
(see section \ref{perc}) 
is such a limit.\footnote%
{In the notation of \cite{class_dyn}, it is the $A\to\infty$ limit of
 the dynamics given by $t_0=1$, $t_n=At^n$, $n=1,2,3,\ldots$.}

%:: percolation is a simple special case
% it's convenient to simulate

Now among these dynamical laws, the one resulting from the choice
$q_n=q^n$ is one of the easiest to work with, both conceptually and
for purposes of computer simulation.  Defined by a single real
parameter $q\in[0,1]$, it is described in more detail in Section
\ref{perc} below.  In \cite{class_dyn}, we referred to it as {\it
transitive percolation} because it can be interpreted in terms of a
random ``turning on'' of nonlocal bonds (with probability $p=1-q$) in
a one-dimensional lattice.  Another thing making it an attractive
special case to work with is the availability in the mathematics
literature of a number of results governing the asymptotic
behavior of posets generated in this manner \cite{mathrefs, AlonEtAl}.

Aside from its convenience, this percolation dynamics, as we will call
it, possesses other distinguishing features, including an underlying
time-reversal invariance and a special relevance to causal set
cosmology, as we 
describe
briefly below.  In this paper, we search for
evidence of a continuum limit of percolation dynamics.

%:: it's T-reversal invariant 
%
%:: [it is an attractor via bounces, hence might be early universe]
% (however this is not limit we study, i think)

One might question whether a continuum limit is even desirable in a
fundamentally discrete theory, but a continuum {\it approximation} in
a suitable regime is certainly necessary if the theory is to reproduce
known physics.  Given this, it seems only a small step to a rigorous
continuum limit, and conversely, the existence of such a limit would
encourage the belief that the theory is capable of yielding continuum
physics with sufficient accuracy.

Perhaps an analogy with kinetic theory can guide us here.  
In quantum gravity, 
the discreteness scale is set, 
presumably, 
by the Planck length $l=(\kappa\hbar)^{1/2}$ (where $\kappa=8\pi{G}$), 
whose vanishing therefore signals a continuum limit.  
In kinetic theory, 
the discreteness scales are set by the mean free path $\lambda$
and the mean free time $\tau$, both of which must go to zero for a
description by partial differential equations to become exact.
Corresponding to these two independent length and time scales are two
``coupling constants'': 
the diffusion constant $D$ and
the speed of sound $c_{\mathrm{sound}}$.  
Just as the value of
the gravitational coupling constant $G\hbar$ reflects 
(presumably) 
the magnitude of the fundamental spacetime discreteness scale, 
so the values of $D$ and $c_{\mathrm{sound}}$ reflect 
the magnitudes of the microscopic parameters $\lambda$ and $\tau$ 
according to the relations
$$
D \sim {\lambda^2 \over \tau} , \quad c_{\mathrm{sound}} \sim
{\lambda \over \tau}
$$
or conversely
$$
   \lambda \sim {D \over c_{\mathrm{sound}}} , \quad 
   \tau \sim {D \over c_{\mathrm{sound}}^2}   \,.
$$
In a continuum limit of kinetic theory,  therefore, we must have 
either $D\to0$ 
or $c_{\mathrm{sound}}\to\infty$.  
In the former case, we can hold
$c_{\mathrm{sound}}$ fixed, but we get a purely mechanical macroscopic
world, without diffusion or viscosity.  In the latter case, we can
hold $D$ fixed, but we get a ``purely diffusive'' world with
mechanical forces propagating at infinite speed.  In each case we get
a well defined --- but defective --- continuum physics, lacking some
features of the true, atomistic world.

If we can trust this analogy, then something very similar must hold in
quantum gravity.  To send $l$ to zero, we must make
either $G$ or $\hbar$ vanish.  In the former case, we would expect to
obtain a quantum world with the metric decoupled from
non-gravitational matter; that is, we would expect to get a theory of
quantum field theory in a purely classical background spacetime
solving the source-free Einstein equations.  In the latter case, we
would expect to obtain classical general relativity.  Thus, there
might be two distinct continuum limits of quantum gravity, each
physically defective in its own way, but nonetheless well defined.

For our purposes in this paper, the important point is that, although
we would not expect quantum gravity to exist as a continuum theory, it
could have limits which do, and one of these limits might be classical
general relativity.  It is thus sensible to inquire whether one of the
classical causal set dynamics we have defined describes classical
spacetimes.  In the following, we make a beginning on this question by
asking whether the special case of ``percolated causal sets'', as we
will call them, admits a continuum limit at all.

Of course, the physical content of any continuum limit we might find
will depend on what we hold fixed in passing to the limit, and this in
turn is intimately linked to how we choose the coarse-graining
procedure that defines the effective macroscopic theory whose
existence the continuum limit signifies.  Obviously, we 
will want 
to send
$N\to\infty$ for any continuum limit, but it is less evident how we
should coarse-grain and what coarse grained parameters we want to hold
fixed in 
taking
the limit.  
Indeed, 
the appropriate choices will depend on 
whether the macroscopic spacetime region we have in mind
is, 
to take some naturally arising examples, 
($i$) a fixed bounded portion of Minkowski space of some dimension, or
($ii$) an entire cycle of a Friedmann universe from initial expansion
to final recollapse, or 
($iii$) an $N$-dependent portion of an unbounded
spacetime $M$ that expands to encompass all of $M$ as $N\to\infty$.  
In the sequel, we will have in mind primarily the 
first of the three examples just
listed.
Without attempting an definitive analysis of the 
coarse-graining question, 
we will simply adopt the simplest definitions
that seem to us to be suited to this example.
More specifically, 
we will coarse-grain by randomly selecting a sub-causal-set of a fixed
number of elements, and we will choose to hold fixed some convenient
invariants of that sub-causal-set, one of which can be
interpreted\footnote% 
{This interpretation is strictly correct only if the causal set forms an
 {\it interval} or ``Alexandrov neighborhood'' within the spacetime.}
as the dimension of the spacetime region it constitutes.
As we will see, the resulting scheme has much in common with the kind of
coarse-graining that goes into the definition of renormalizability in
quantum field theory.  For this reason, we believe it can serve also as
an instructive ``laboratory'' in which this concept, 
and related concepts like 
``running coupling constant'' and
``non-trivial fixed point'', 
%% renormalization group scaling 
can be considered from a fresh perspective.

In the remaining sections of this paper we: define transitive
percolation dynamics more precisely; specify the coarse-graining
procedure we have used; report on the simulations we have run looking
for a continuum limit in the sense thereby defined; and offer some
concluding comments.

\subsection{Definitions used in the sequel}

Causal set theory postulates that spacetime, at its most fundamental
level, is discrete, and that its macroscopic geometrical properties
reflect a deep structure which is purely order theoretic in nature.
This deep structure is taken to be a partial order
and called a causal set (or ``causet'' for short). 
For an introduction to causal set theory, see
\cite{bombelli,causets2,reid,causets1}.  In this section, we merely
recall some definitions which we will be using in the sequel.

A {\it (partial) order} or {\it poset} is a set $S$ endowed with a
relation $\prec$ which is:
\bee
 \mbox{transitive} \qquad & 
        \forall x,y,z \in S \qquad 
        x \prec y 
        \  \mbox{and} \  
        y \prec z 
	\quad \Rightarrow \quad 
	x \prec z \\ \
 \mbox{acyclic} \qquad & 
        \forall x,y \in S \qquad 
        x \prec y \Rightarrow y \not\prec x \\ 
 \mbox{irreflexive} \qquad &  
        \forall x \in S \qquad  
        x \not\prec x 
\eee 
(Irreflexivity is 
merely a 
convention; with it, acyclicity is
actually redundant.)  For example, the events of Minkowski space (in
any dimension) form a poset whose order relation is the usual causal
order.  In an order $S$, the \emph{interval} $\interval(x,y)$ is
defined to be
\be
          \interval(x,y) = \{ z \in S | x \prec z \prec y \} \ .  
\ee 
An order is said to be \emph{locally finite} if all its intervals are
finite (have finite cardinality).  A \emph{causal set} is a locally
finite order.

\begin{figure}[htbp]
\label{names}
\center
\scalebox{.7}{\includegraphics{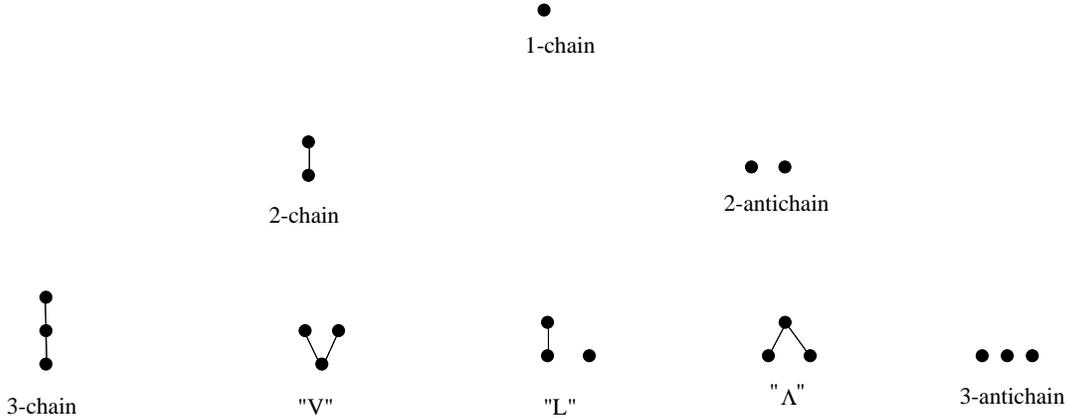}}
\caption{Names for small causets}
\end{figure}

It will be helpful to have names for some small causal sets.  Figure
\ref{names} provides such names for the causal sets with three or
fewer elements.

%: The dynamics of transitive percolation 

\section{The dynamics of transitive percolation}
\label{perc}

Regarded as a sequential growth dynamics of the sort derived in
\cite{class_dyn}, transitive percolation is described by one free
parameter $q$ such that $q_n = q^n$.  This is equivalent (at stage $N$
of the growth process) to using the following ``percolation''
algorithm to generate a random causet.

\begin{enumerate}
\item Start with $N$ elements labeled $0, 1, 2, \ldots, N-1$.

\item With a fixed probability $p$ $(=1-q)$, introduce a relation
$i\prec j$ between every pair of elements  labeled $i$ and $j$, where $i
\in \{0\cdots N-2\}$ and $j \in \{i+1\cdots N-1\}$.

\item Form the transitive closure of these relations  
(e.g. if $2\prec5$ and $5\prec8$ then enforce that $2\prec8$.)

\end{enumerate}
Given the simplicity of this dynamical model, both conceptually and
from an algorithmic standpoint, it offers a ``stepping stone''
allowing us to look into some general features of causal set dynamics.
(The name ``percolation'' comes from thinking of a relation 
$i\prec j$ as a ``bond'' or ``channel'' between $i$ and $j$.)

%% \subsection{Originary transitive percolation}

There exists another model which is very similar to transitive
percolation, called ``originary transitive percolation''.  The rule
for randomly generating a causet is the same as for transitive
percolation, except that each new element is required to be related to
at least one existing element.  
Algorithmically, we generate
potential elements one by one, exactly as for plain percolation, but
discard any such element which would be unrelated to all previous
elements.  
Causets formed with this dynamics always have a single
minimal element, an ``origin''.

Recent work by Dou \cite{dou} suggests that originary percolation
might have an important role to play in cosmology.  Notice first that,
if a given cosmological ``cycle'' ends with the causet collapsing down
to a single element, then the ensuing re-expansion is necessarily
given by an originary causet.  Now, in the limited context of
percolation dynamics, Alon {\it et al.} have proved rigorously
\cite{AlonEtAl} that such cosmological ``bounces'' (which they call
{\it posts}) occur with probability 1 (if $p>0$), from which it follows
that there are infinitely many cosmological cycles, each cycle but the
first having the dynamics of originary percolation.  For more general
choices of the dynamical parameters $q_n$ of \cite{class_dyn}, posts
can again occur, but now the $q_n$ take on new effective values in
each cycle, related to the old ones by the action of a sort of
``cosmological renormalization group''; and Dou \cite{dou} has found
evidence that originary percolation is a ``stable fixed point'' of
this action, meaning that the universe would tend to evolve toward
this behavior, no matter what dynamics it began with.

It would thus be of interest to investigate the continuum limit of
originary percolation as well as plain percolation.  In the present
paper, however, we limit ourselves to the latter type, 
which we believe is more appropriate 
(albeit not fully appropriate for reasons discussed in the conclusion)
in the context of spacetime regions of sub-cosmological scale.

%: The Critical point at $p=0$, $N=\infty$

\def\Nc2{{N \choose 2}}

\section{The critical point at $p=0$, $N=\infty$}
In the previous section we have introduced a model of random causets,
which depends on two parameters, $p\in [0,1]$ and
$N\in\NaturalNumbers$.  For a given $p$, the model defines a
probability distribution on the set of $N$-element causets.\footnote%
{Strictly speaking this distribution has gauge-invariant meaning only in
 the limit $N\to\infty$ ($p$ fixed); for it is only insofar as the growth
 process ``runs to completion'' that generally covariant questions can
 be asked.  Notice that this limit is inherent in causal set dynamics
 itself, and has nothing to do with the continuum limit we are concerned
 with herein, which sends $p$ to zero as $N\to\infty$.}
For $p=0$, the only causet with nonzero probability, obviously, is the
$N$-antichain.  Now let $p>0$.  With a little thought, one can
convince oneself that for $N\rightarrow\infty$, the causet will look
very much like a chain.  Indeed it has been proved \cite{bb} (see also
\cite{schulman}) that, as $N\rightarrow\infty$ with $p$ fixed at some
(arbitrarily small) positive number, $r\rightarrow1$ in probability,
where
\be 
      r \equiv \frac{R}{N(N-1)/2} = { R \over \Nc2 } \,, 
\ee
$R$ being the number of relations in the causet,
i.e. the number of pairs of causet elements $x$, $y$ 
such that $x\prec{y}$ or $y\prec{x}$.
Note that the $N$-chain has the greatest possible number $\Nc2$ of
relations, so $r\rightarrow1$ gives a precise meaning to ``looking
like a chain''.  We call $r$ the \emph{ordering fraction} of the
causal set, following \cite{myrheim}.

We see that for $N\rightarrow\infty$, there is a change in the
qualitative nature of the causet as $p$ varies away from zero, and the
point $p=0, N=\infty$ (or $p=1/N=0$) is in this sense a
critical point of the model.  It is the behavior of the model near this
critical point which will concern us in this paper.

%: Coarse graining

\section{Coarse graining}
\label{coarse_graining}
An advantageous feature of causal sets is that there exists for them a
simple yet precise notion of coarse graining.  A coarse grained
approximation to a causet $C$ can be formed by selecting a sub-causet
$C'$ at random, with equal selection probability for each element, and
with the causal order of $C'$ inherited directly from that of $C$
(i.e. $x\prec{}y$ in $C'$ if and only if $x\prec{}y$ in $C$.)

\begin{figure}[htbp]
\center
\scalebox{.5}{\includegraphics{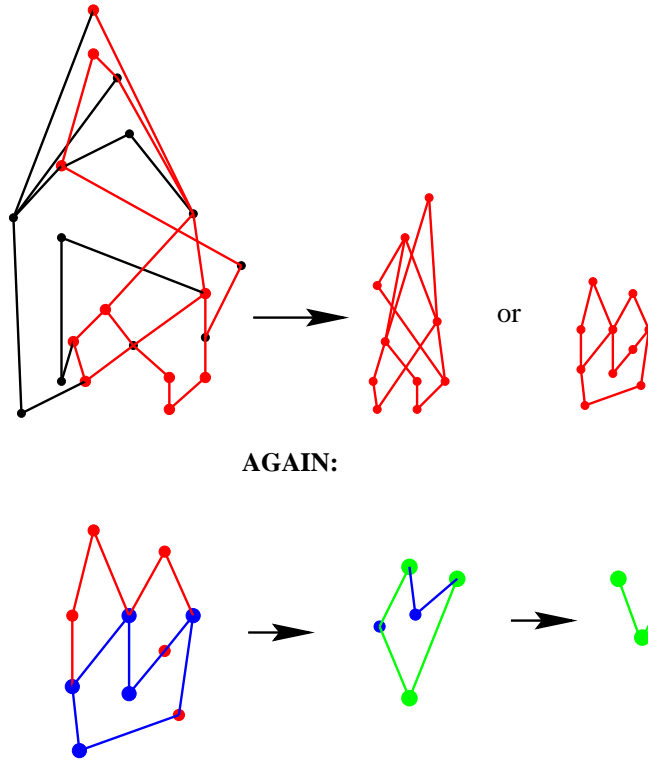}} % .6
\caption{Three successive coarse grainings of a 20-element causet}
\label{coarse_grain}
\end{figure}

For example, let us start with the $20$ element causet $C$ shown in 
Figure \ref{coarse_grain}.
%% RDS: I don't know why the number came out wrong, namely as 4. 
% DPR: It should work now.
(which was percolated using $p=0.25$), and successively coarse grain 
it down to causets of 10, 5 and 3 elements. 
We see that, at the largest scale shown 
(i.e. the smallest number of remaining elements), 
$C$ has coarse-grained 
in this instance to the 3-element ``V'' causet.  
Of course, 
coarse graining itself is a random process, 
so from a single causet of $N$ elements, 
it gives us in general, 
not another single causet, 
but a probability distribution 
on the causets of $m<N$ elements.

A noteworthy feature of this definition of coarse graining, which in
some ways is similar to what is often called ``decimation'' in the
context of spin systems, is the {\it random} selection of a subset.
In the absence of any background lattice structure to refer to, no
other possibility for selecting a sub-causet is evident.  
Random selection is also 
recommended 
strongly by considerations of Lorentz
invariance \cite{daughton}.  The fact that a coarse grained causet is
automatically another causet will make it easy for us to formulate
precise notions of continuum limit, running of the coupling constant
$p$, etc.  In this respect, we believe that this model combines
precision with novelty in such a manner as to furnish an instructive
illustration of concepts related to renormalizability, independently
of its application to quantum gravity.  We remark in this connection,
that transitive percolation is readily embedded in a
``two-temperature'' statistical mechanics model, and as such, happens
also to be exactly soluble in the sense that the partition function
can be computed exactly \cite{AChR,scaling}.

%: The large scale effective theory

\section{The large scale effective theory}

In section \ref{perc} we described a ``microscopic'' dynamics for
causal sets (that of transitive percolation) and in section
\ref{coarse_graining} we defined a precise notion of coarse graining
(that of random selection of a sub-causal-set).  
On this basis,
we can produce an effective ``macroscopic'' dynamics by imagining that
a causet 
$C$
is first percolated with $N$ elements and then coarse-grained
down to $m<N$ elements.  This two-step process constitutes an
effective 
random
procedure for generating $m$ element causets depending (in
addition to $m$) on the parameters $N$ and $p$.  In causal set theory,
number of elements corresponds to spacetime volume, so we can
interpret $N/m$ as the factor by which the ``observation scale'' has
been increased by the coarse graining.  
If, then, $V_0$ is the macroscopic volume of the spacetime region
constituted by our causet, and if we take $V_0$ to be fixed as
$N\to\infty$, then our procedure for generating causets of $m$ elements
provides the effective dynamics at volume-scale $V_0/m$ (i.e. length
scale $(V_0/m)^{1/d}$ for a spacetime of dimension $d$).

What does it mean for our effective theory to have a continuum limit
in this context?  Our stochastic microscopic dynamics gives, for each
choice of $p$, a probability distribution on the set of causal sets
$C$
with $N$ elements, and by choosing $m$, we determine at which scale we
wish to examine the corresponding effective theory.  This effective
theory is itself just a probability distribution $f_m$ on the set of
$m$-element causets, and so our dynamics will have a well defined
continuum limit if there exists, as $N\to\infty$, a trajectory
$p=p(N)$ along which 
the corresponding probability distributions $f_m$
on coarse grained causets 
approach fixed limiting distributions
$f_m^\infty$
for all $m$.
The limiting theory in this sense is then a sequence of effective
theories, one for each $m$, all fitting together consistently. 
(Thanks to the associative (semi-group)
character of our coarse-graining procedure, the existence of a
limiting distribution for any given $m$ implies its existence for all
smaller $m$.
Thus it suffices that a limiting distribution $f_m$ exist for $m$
arbitrarily large.)  
In general there will exist not just a single such
trajectory $p=p(N)$, but a one-parameter family of them (corresponding
to the one real parameter $p$ that characterizes the microscopic
dynamics at any fixed $N$), and one may 
wonder whether
all the
trajectories will take on the same asymptotic form as they approach
the critical point $p=1/N=0$.

Consider first the simplest nontrivial case, $m=2$.  Since there are
only two causal sets of size two, the 2-chain and the 2-antichain, the
distribution $f_2$ that gives the ``large scale physics'' in this case
is described by a single number which we can take to be $f_2(\twoch)$,
the probability of obtaining a 2-chain rather than a 2-antichain.
(The other probability, $f_2(\twoach)$, is of course not independent,
since classical probabilities must add up to unity.)

Interestingly enough, the number $f_2(\twoch)$ has a direct physical
interpretation in terms of the Myrheim-Meyer dimension of the
fine-grained causet $C$.  
Indeed, it is easy to see that $f_2(\twoch)$ is nothing but
the expectation value of what
we called above the ``ordering fraction'' of $C$.  
But the ordering fraction, in turn, determines the 
Myrheim-Meyer dimension $d$ that indicates the dimension of
the Minkowski spacetime $\Minkowski^d$ (if any) in which $C$ would
embed faithfully as an interval \cite{meyer, myrheim}.  
Thus, by coarse graining down to two elements, we are effectively
measuring a certain kind of spacetime dimensionality of $C$.
In practice, we would not expect $C$ to embed faithfully without some
degree of coarse-graining, but the original $r$ would still provide a
good dimension estimate since it is, on average, coarse-graining invariant.

%% [[RDS: above rewritten to fix problem you pointed out ]] 

As we begin to consider 
coarse-graining to
sizes $m>2$, 
the degree of complication grows
rapidly, simply because the number of partial orders defined on $m$
elements grows rapidly with $m$.  For $m=3$ there are five possible
causal sets: \threech, \V, \Lcauset, \wedge, and \threeach.  Thus the
effective dynamics at this ``scale'' is given by five probabilities
(so four free parameters).  For $m=4$ there are sixteen
probabilities, for $m=5$ there are sixty three, and for $m=6$, 7 and
8, the number of probabilities is respectively 318, 2045 and 16999.

%: Simulations

\section{Evidence from simulations}

In this section, we report on some computer simulations that address
directly the question whether transitive percolation possesses a
continuum limit in the sense defined above.  In a subsequent paper, we
will report on simulations addressing the subsidiary question of a
possible scaling behavior in the continuum limit.

In order that a continuum limit exist, it must be possible to choose a
trajectory for $p$ as a function of $N$ so that the resulting
coarse-grained probability distributions, $f_1$, $f_2$, $f_3$, \dots,
have well defined limits as $N\to\infty$.  To study this question
numerically, one can simulate transitive percolation using the
algorithm described in Section \ref{perc}, while choosing $p$ so as to
hold constant (say) the $m=2$ distribution $f_2$ ($f_1$ being
trivial).  Because of the way transitive percolation is defined, it is
intuitively obvious that $p$ can be chosen to achieve this, and that
in doing so, one leaves $p$ with no further freedom.  The decisive
question then is whether, along the trajectory thereby defined, the
higher distribution functions, $f_3$, $f_4$, etc.  all approach
nontrivial limits.

As we have already mentioned, holding $f_2$ fixed is the same thing as
holding fixed the expectation value $\E{r}$ of ordering fraction
$r=R/{{N}\choose{2}}$.  To see in more detail why this is so, consider
the coarse-graining 
%% process 
that takes us from the original causet
$C_N$ of $N$ elements to a causet $C_2$ of two elements.  
Since coarse-graining is just random selection, the probability
$f_2(\twoch)$ that $C_2$ turns out to be a 2-chain is just the
probability that two elements of $C_N$ selected at random form a
2-chain rather than a 2-antichain.  
In other words, it is just the
probability that two elements of $C_N$ selected at random are 
causally related.  
Plainly, this is the same as the {\it fraction} of pairs of
elements of $C_N$ such that the two members of the pair form a
relation $x\prec{y}$ or $y\prec{x}$.  Therefore, the ordering fraction
$r$ equals the probability of getting a 2-chain when coarse graining
$C_N$ down to two elements; and $f_2(\twoch)=\E{r}$, as claimed.

This reasoning illustrates, 
in fact, 
how one can 
in principle 
determine 
any one of the
distributions $f_m$ 
by answering the question,
``What is the probability of getting this particular $m$-element
causet from this particular $N$-element causet if you coarse grain
down to $m$ elements?''  To compute the answer to such a question
starting with any given causet $C_N$, one examines every possible
combination of $m$ elements, counts the number of times that the
combination forms the particular causet being looked for, and divides
the total by ${N \choose m}$.  The ensemble mean of the resulting {\it
abundance}, as we will refer to it, is then $f_m(\xi)$, where $\xi$ is
the causet being looked for.  In practice, of course, we would normally use
a more efficient counting algorithm than simply examining individually
all ${N \choose m}$ subsets of $C_N$.

\subsection{Histograms of 2-chain and 4-chain abundances}

\begin{figure}[htbp]
\center
\scalebox{.73}{ % .74 too big
\rotatebox{-90}{
\includegraphics{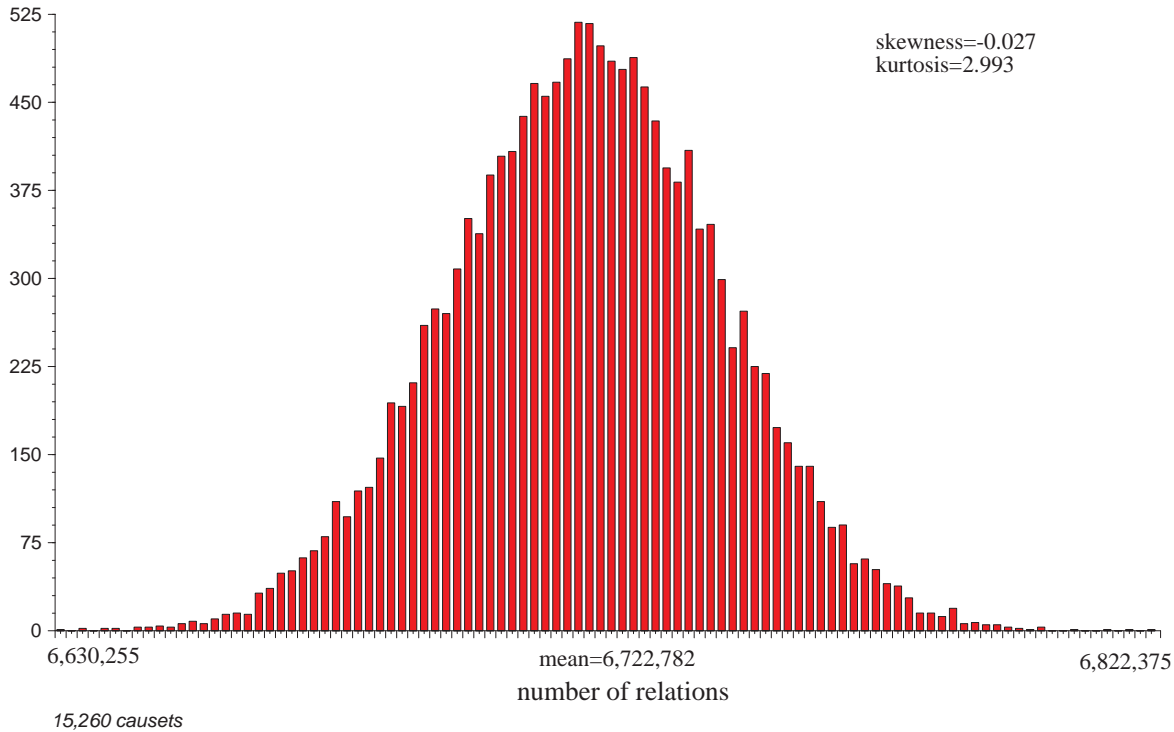} }}
\label{hist2ch}
\caption{Distribution of number of relations for $N=4096$, $p=0.01155$}
\end{figure}

\begin{figure}[htbp]
\center \scalebox{.7}{ \rotatebox{-90}{
\includegraphics{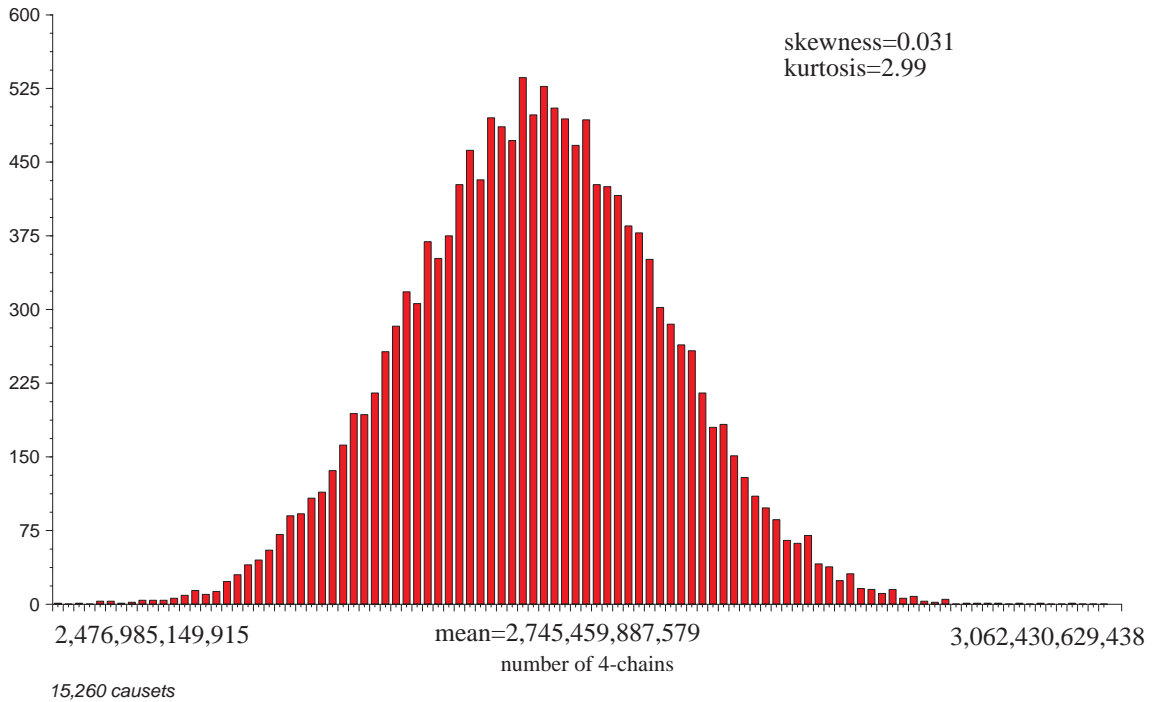} }}
\caption{Distribution of number of 4-chains for $N=4096$, $p=0.01155$}
\label{hist4ch}
\end{figure}

As explained in the previous subsection, the main computational
problem, once the random causet has been generated, is determining the
number of subcausets of different sizes and types.  To get a feel for
how some of the resulting ``abundances'' are distributed, we start by
presenting a couple of histograms.
Figure \ref{hist2ch} shows the number $R$ of relations
obtained from a simulation in which 15,260 causal sets were generated by
transitive percolation with $p=0.01155$, $N=4096$.
Visually, the distribution is Gaussian, in agreement with
the fact that its ``kurtosis''
\be
  \overline{ \left(x - \bar{x} \right)^4 }   \, \bigg/ \       
  {\overline{ \left(x-\bar{x}\right)^2 }\,}^2
\ee
of 2.993 is very nearly equal to its Gaussian value of 3 
(the over-bar denotes sample mean).  
In these simulations, $p$ was chosen so that the number of 3-chains 
was equal on average to 
half the total number possible, 
i.e. the ``abundance of 3-chains'',
$\mbox{(number of 3-chains)}/{N \choose 3}$, 
%% $\frac{\mbox{number of 3-chains}}{{N \choose 3}}$, 
was
equal to $1/2$ on
average.  The picture is qualitatively identical if one counts
4-chains rather than 2-chains, as exhibited in Fig. \ref{hist4ch}.

(One may wonder whether it was to be expected that these distributions
 would appear to be so normal.  If the variable in question, here the
 number of 2-chains $R$ or the number of 4-chains ($C_4$, say), 
 can
 be expressed as a sum of independent random variables, then the
 central limit theorem 
 provides an explanation.  So consider the
 variables $x_{ij}$ which are 1 if $i \prec j$ and zero otherwise.
 Then 
 $R$ is easily expressed as a sum of these variables:
\be
        R = \sum_{i<j} x_{ij}
\ee
However, 
the $x_{ij}$
are
not independent, 
due to transitivity.  
Apparently,  this dependence 
is not large enough to interfere much with the
normality of their sum.  The number of 4-chains $C_4$ can be expressed
in a similar manner
\be
    C_4 = \sum_{i<j<k<l} x_{ij} x_{jk} x_{kl}\,.
\ee
and similar remarks apply.)

Let us mention that for values of $p$ sufficiently close to 0 or 1,
these distributions will appear skew.  This occurs simply because the
numbers under consideration (e.g. the number of $m$-chains) are
bounded between zero and $N \choose m$ and must deviate from normality
if their mean gets too close to a boundary relative to the size of
their standard deviation.  Whenever we draw an error bar in the
following, we will ignore any deviation from normality in the
corresponding distribution.

Notice incidentally that the total number of 4-chains possible is
${4096\choose4}=11,710,951,848,960$.  Consequently, the mean 4-chain
abundance\footnote
{From this point on we will usually write simply ``abundance'', in
place of ``mean abundance'', assuming the average is obvious from context.}
in our simulation is only
$\frac{2,745,459,887,579}{11,710,951,848,960}=0.234$, a considerably smaller
value than the 2-chain abundance of $r=\frac{6,722,782}{{4096 \choose
2}}=0.802$.  This was to be expected, considering that the 2-chain is
one of only two possible causets of its size, while the 4-chain is one
out 16 possibilities.  (Notice also that 4-chains are necessarily less
probable than 2-chains, because every coarse-graining of a 4-chain is
a 2-chain, whereas the 2-chain can come from every 4-element causet
save the 4-antichain.)

\subsection{Trajectories of $p$ versus $N$}

The 
question we 
are exploring
is 
whether
there exist, for $N\to\infty$, trajectories $p=p(N)$ along which the
mean abundances of all finite causets tend to definite limits.
To seek such trajectories numerically, we will select some finite
``reference causet'' and determine, for a range of $N$, those values
of $p$ which maintain its abundance at some target value.  If a
continuum limit does exist, then it should not matter in the end which
causet we select as our reference, since any other choice (together
with 
a matching
choice of target abundance) should produce the
same trajectory asymptotically.  We would also anticipate that all the
trajectories would behave similarly for large $N$, and that, in
particular, either all would lead to continuum limits or all would
not.  In principle it could happen that only a certain subset led to
continuum limits, but we know of no reason to expect such an
eventuality.
In the simulations reported here, we have 
chosen as our reference
causets the 2-, 3- and 5-chains.  We have computed six trajectories,
holding the 2-chain abundance fixed at 1/2, 1/3, and 1/10, the 3-chain
abundance fixed at 1/2 and .0814837, and the 5-chain abundance fixed
at 1/2.  For $N$, we have used as large a range as our computers would
allow.

\begin{figure}[htbp!]
\center
\scalebox{1.22}{
\includegraphics{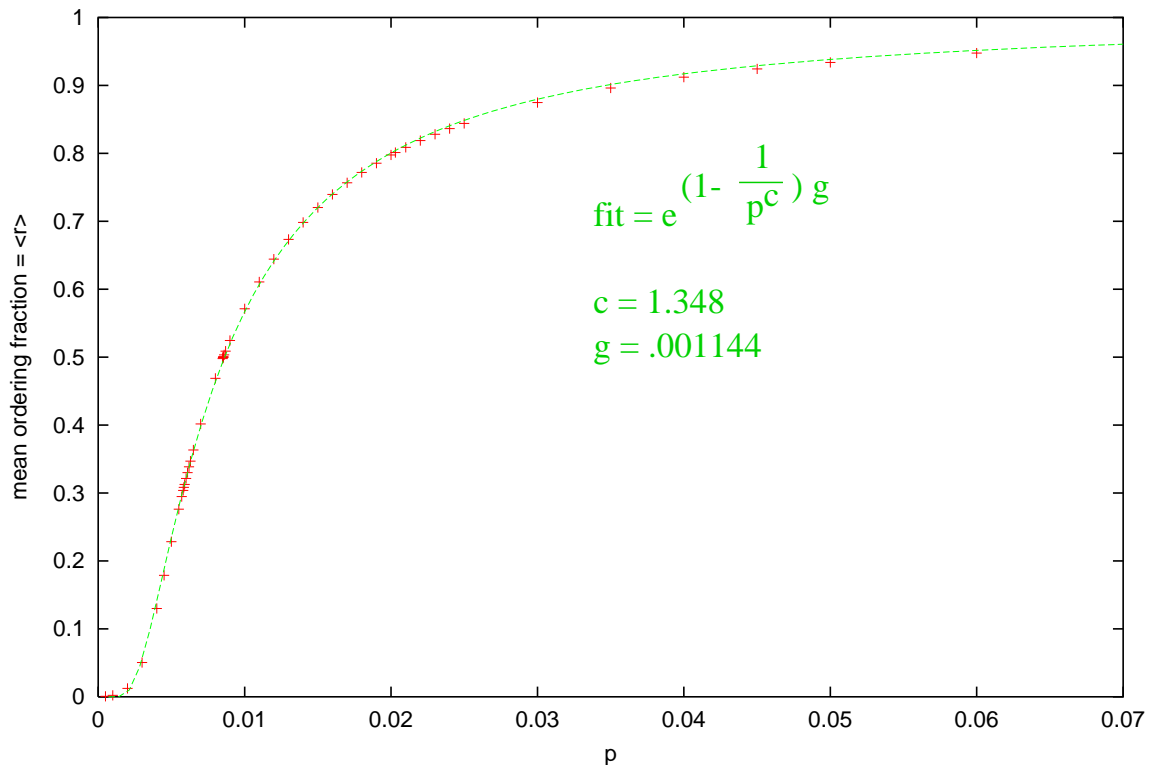} }
\caption{Ordering fractions as a function of $p$ for $N = 2048$}
\label{rvsp}
\end{figure}

Before discussing the trajectories as such, let us have a look at how
the mean 2-chain abundance $\E{r}$ (i.e. the mean ordering fraction)
varies with $p$ for a fixed $N$ of 2048, as exhibited in Figure \ref{rvsp}.  
(Vertical error bars are displayed in the figure but are
so small that they just look like horizontal lines.  The plotted
points were obtained from an exact expression for the ensemble average
$\E{r}$, so the errors come only from floating point roundoff.  The
fitting function used in Figure \ref{rvsp} will be discussed in a
subsequent paper \cite{scaling}, where we examine scaling behavior;
see also \cite{mathrefs}.)
As one can see, $\E{r}$ starts at 0 for $p=0$, rises rapidly to near 1
and then asymptotes to 1 at $p=1$ (not shown).  Of course, it was
evident a priori that $\E{r}$ would increase monotonically from 0 to 1
as $p$ varied between these same two values, but it is perhaps
noteworthy that its graph betrays no sign of discontinuity or
non-analyticity (no sign of a ``phase transition'').  To this extent,
it strengthens the expectation that the trajectories we find will all
share the same qualitative behavior as $N\to\infty$.

\begin{figure}[htbp]
\center
\scalebox{.78}{
\includegraphics{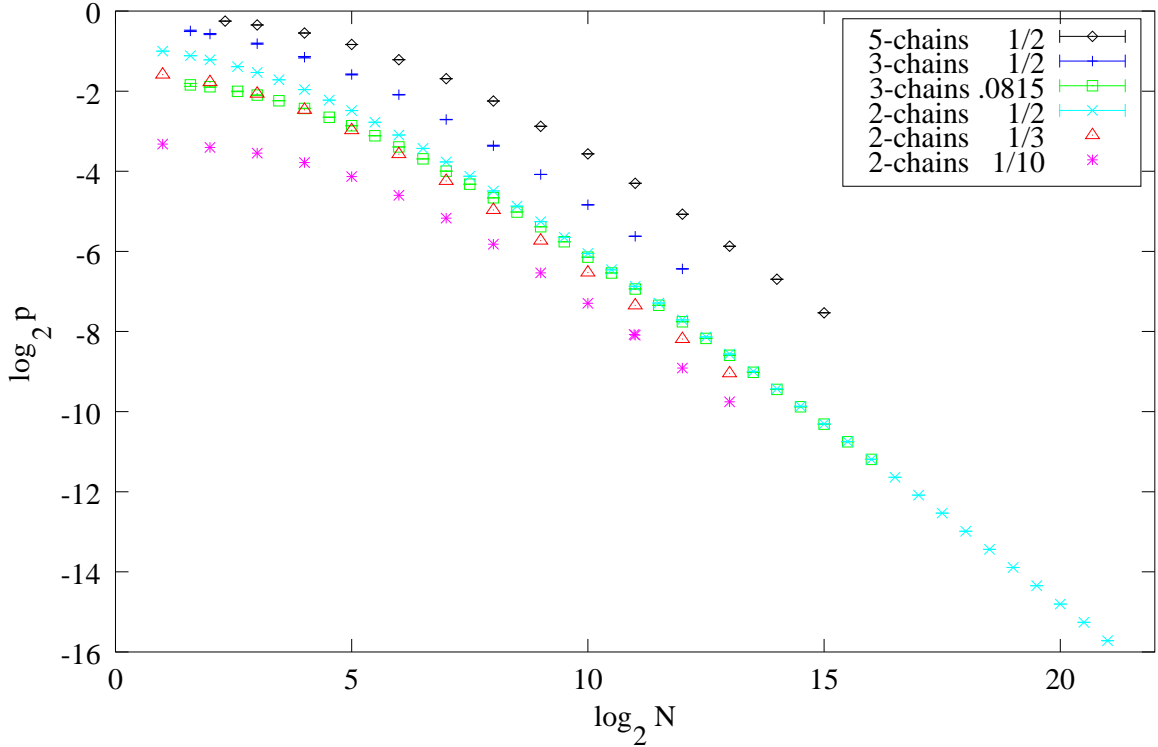} }
\caption%
{Flow of the ``coupling constant'' $p$ as $N\to\infty$ (six trajectories)}
\label{6traj}
\end{figure}

The six trajectories we have simulated are depicted in
Fig. \ref{6traj}.\footnote%
{Notice that the error bars are shown rotated in the legend.  This will
 be the case for all subsequent legends as well.}
A higher
abundance of $m$-chains for 
fixed
$m$ leads to a
trajectory
with
higher $p$.  
Also note that, as observed above, the longer chains
require larger values of $p$ to attain the same mean abundance, hence
a choice of mean abundance = 1/2  
corresponds in each case to a
different trajectory.
The trajectories with $\E{r}$ held to 
lower values
are 
``higher dimensional'' 
in the sense that
$\E{r}=1/2$ corresponds to a Myrheim-Meyer
dimension of 2, while $\E{r}=1/10$ corresponds to a Myrheim-Meyer
dimension of 4.
Observe that the
plots
give the impression of becoming
straight lines with a common slope at large $N$.  
This tends to
corroborate
the
expectation that they will exhibit some form of
scaling with a common exponent, a behavior reminiscent of that found
with continuum limits in many other contexts.
This is further suggested by the fact that two distinct trajectories
($f_2(\twoch)=1/2$ and $f_3(\threech)=.0814837$), obtained by holding
different abundances fixed, seem to converge for large $N$.
% close, slopes become equal

%% RDS: change caption on figure {critpoint} to: 
%%  ``Flow toward the critical point'' (no `s' after toward)

\begin{figure}[htbp]
\center
\scalebox{.80}{%1.15
\includegraphics{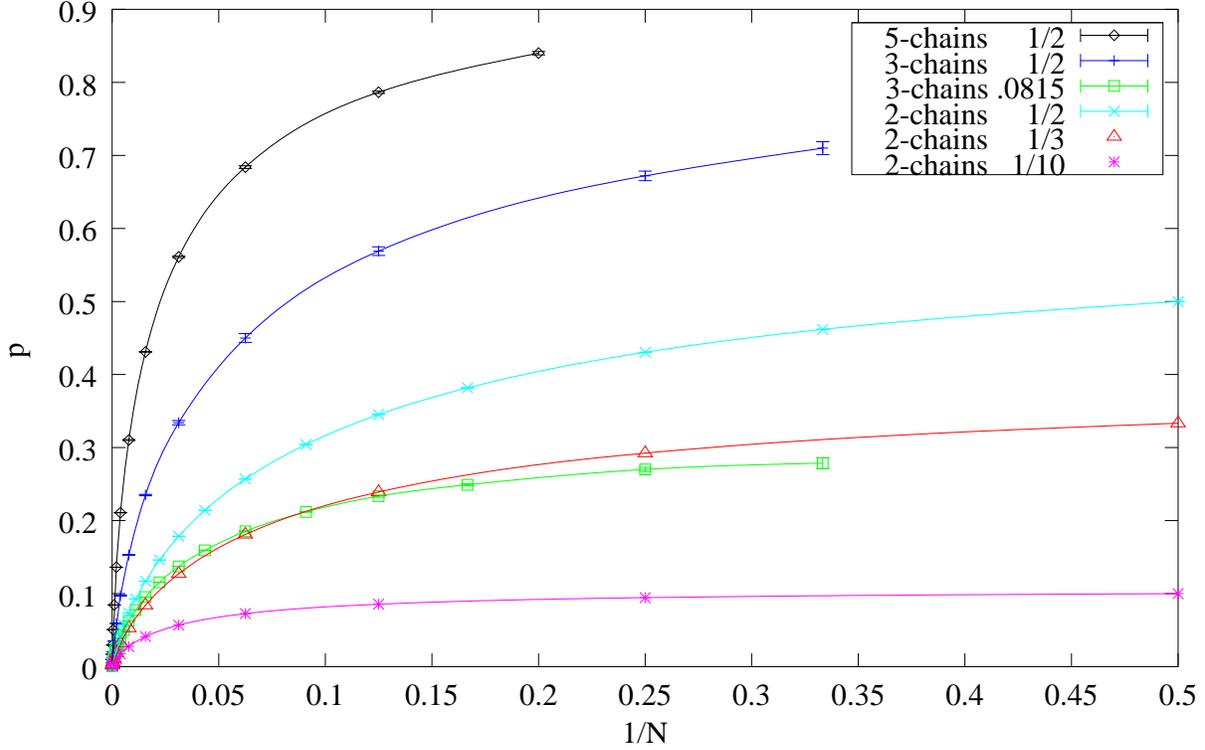} }
\caption{Six trajectories approaching the critical point at $p=0$, $N=\infty$}
\label{critpoint}
\end{figure}

By taking the abscissa to be $1/N$ rather than $\log_2 N$, 
we can bring the critical point to the origin,
as in
Fig. \ref{critpoint}.  The lines which pass through the data points
there are just splines drawn to aid the eye in following the
trajectories.
Note that the curves tend to asymptote to the $p$-axis, suggesting
that $p$ falls off more slowly than $1/N$.  
This suggestion is corroborated by 
more detailed analysis of the scaling behavior of these trajectories, 
as will be discussed in \cite{scaling}.
% logarithmic corrections?

\subsection{Flow of the coarse-grained theory along a trajectory}

We come finally to a direct test of whether the coarse-grained theory
converges to a limit as $N\to\infty$.  Independently of scaling or any
other indicator, this is by definition the criterion for a continuum
limit to exist.  We have examined this question by means of
simulations conducted for five of the six trajectories mentioned
above.  In each simulation we proceeded as follows.  For each chosen
$N$, we experimentally found a $p$ sufficiently close to the desired
trajectory.  Having determined $p$, we then generated a large number
of causets by the percolation algorithm described in
Section~\ref{perc}.  (The number generated varied from 64 to 40,000.)
For each such random causet, we computed the abundances of the
different $m$-element (sub)causets under consideration (2-chain, 3-chain,
3-antichain, etc), and we combined the results to obtain the mean
abundances we have plotted here, together with their standard errors.
(The errors shown do not include any contribution from the slight
inaccuracy in the value of $p$ used.  Except for the 3- and 5-chain
trajectories these errors are negligibly small.)

To compute the abundances of the 2-, 3-, and 4-orders for a given
causet, 
we randomly sampled its four-element subcausets, 
counting the number of times each of the sixteen possible
4-orders arose, and  dividing each of these counts by the number of
samples taken to get the corresponding abundance.
As an aid in identifying
to which 4-order 
a sampled subcauset belonged
we used the following
invariant, 
which distinguishes all of the sixteen 4-orders, 
save two pairs.  
$$
      I(S) = \prod\limits_{x\in S} \left( 2 + |\past(x)| \right)
$$
Here, $\past(x) = \{y\in S | y \prec x\}$ 
is the exclusive past of the element $x$ and $|\past(x)|$ is its cardinality.
Thus, we associate to each element of the causet, 
a number which is two more than the cardinality of its exclusive past, 
and we form the product of these numbers 
(four, in this case) 
to get our invariant.  
(For example, this invariant is 90 for the ``diamond'' poset, \diamond.)
  
\begin{figure}[htbp]
\center
\scalebox{.78}{
\includegraphics{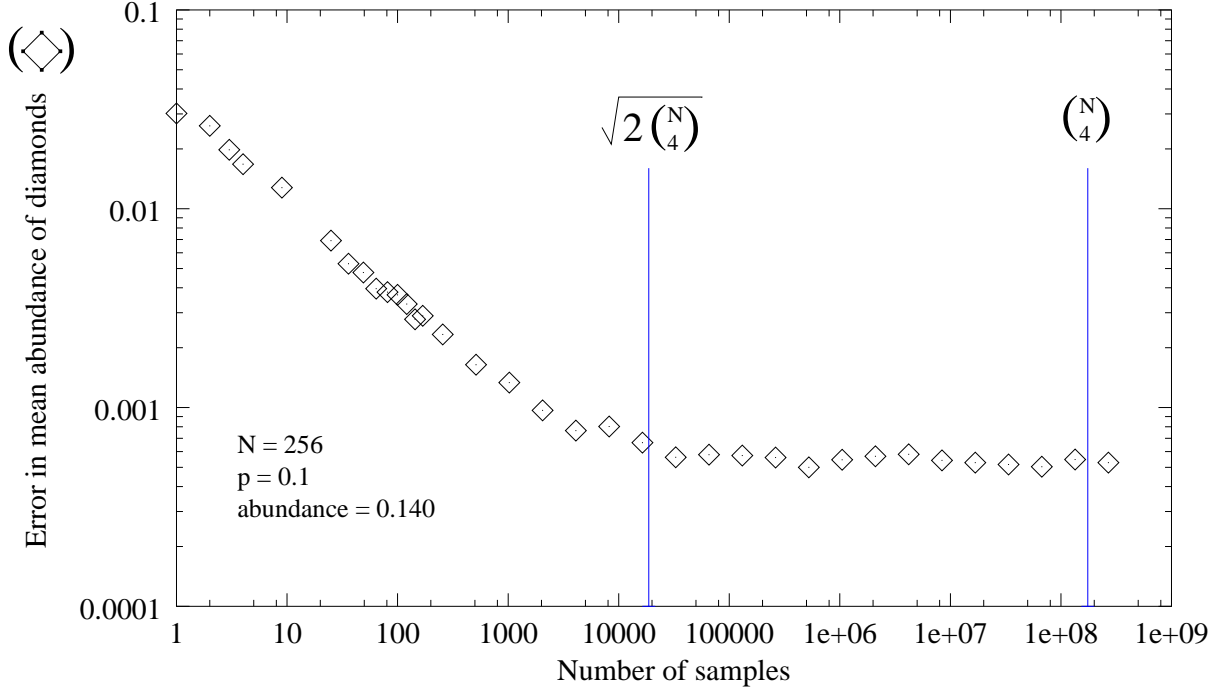} }
\caption{Reduction of error in estimated diamond abundance with
         increasing number of samples} 
\label{sampling_error}
\end{figure}

The number of samples taken
from
an $N$
element causet was chosen to be $\sqrt{2{N \choose 4}}$, 
on the grounds that
the
probability to get the same four element subset twice becomes
appreciable with more than this many samples.  
Numerical tests
confirmed that 
this rule of thumb
tends to minimize 
the sampling error, 
as seen in Figure \ref{sampling_error}.

Once 
one has
the abundances of all the 4-orders, the abundances of the
smaller causets can be found by further coarse graining.  By
explicitly carrying out this coarse graining, one 
easily deduces
the following relationships:
\bee
f_3(\threech) & = & f_4(\fourch) + \half\left(f_4(\Y)+f_4(\iY)\right) 
	+ \quart f_4(\threecho) + \quart\left(f_4(\new)+f_4(\inu)\right) 
	+ \half f_4(\diamond)\\
f_3(\V) & = & \half\, f_4(\Y) + \half f_4(\new) + \quart f_4(\diamond) +
	\thquart f_4(\flower) + \quart f_4(\Vo) + \quart f_4(\N) 
	+ \half f_4(\bowtie)\\
f_3(\Lcauset) & = & \thquart f_4(\threecho) +
	\quart\left(f_4(\new)+f_4(\inu)\right) +
	\half\left(f_4(\Vo)+f_4(\wedgeo)\right) + f_4(\pie) + \half f_4(\Lo) +
	\half f_4(\N)\\
f_3(\wedge) & = & \half\, f_4(\iY) + \half f_4(\inu) + \quart f_4(\diamond) +
	\thquart f_4(\iflower) + \quart f_4(\wedgeo) + \quart f_4(\N) 
	+ \half f_4(\bowtie)\\
f_3(\threeach) & = & \quart\left(f_4(\flower)+f_4(\iflower)\right) 
	+ \quart\left(f_4(\Vo)+f_4(\wedgeo)\right) + \half f_4(\Lo) 
	+ f_4(\fourach)\\
f_2(\twoch) & = & f_3(\threech) +
	\frac{2}{3}\left(f_3(\V)+f_3(\wedge)\right) 
	+ \frac{1}{3} f_3(\Lcauset)\\
f_2(\twoach) & = & 1 - f_2(\twoch)
\eee
In the first six equations, 
the coefficient before each term on the right is the 
fraction
of
coarse-grainings 
of that causet
which yield the causet on the left.

\begin{figure}[htbp]
\center
\scalebox{.70}{
\includegraphics{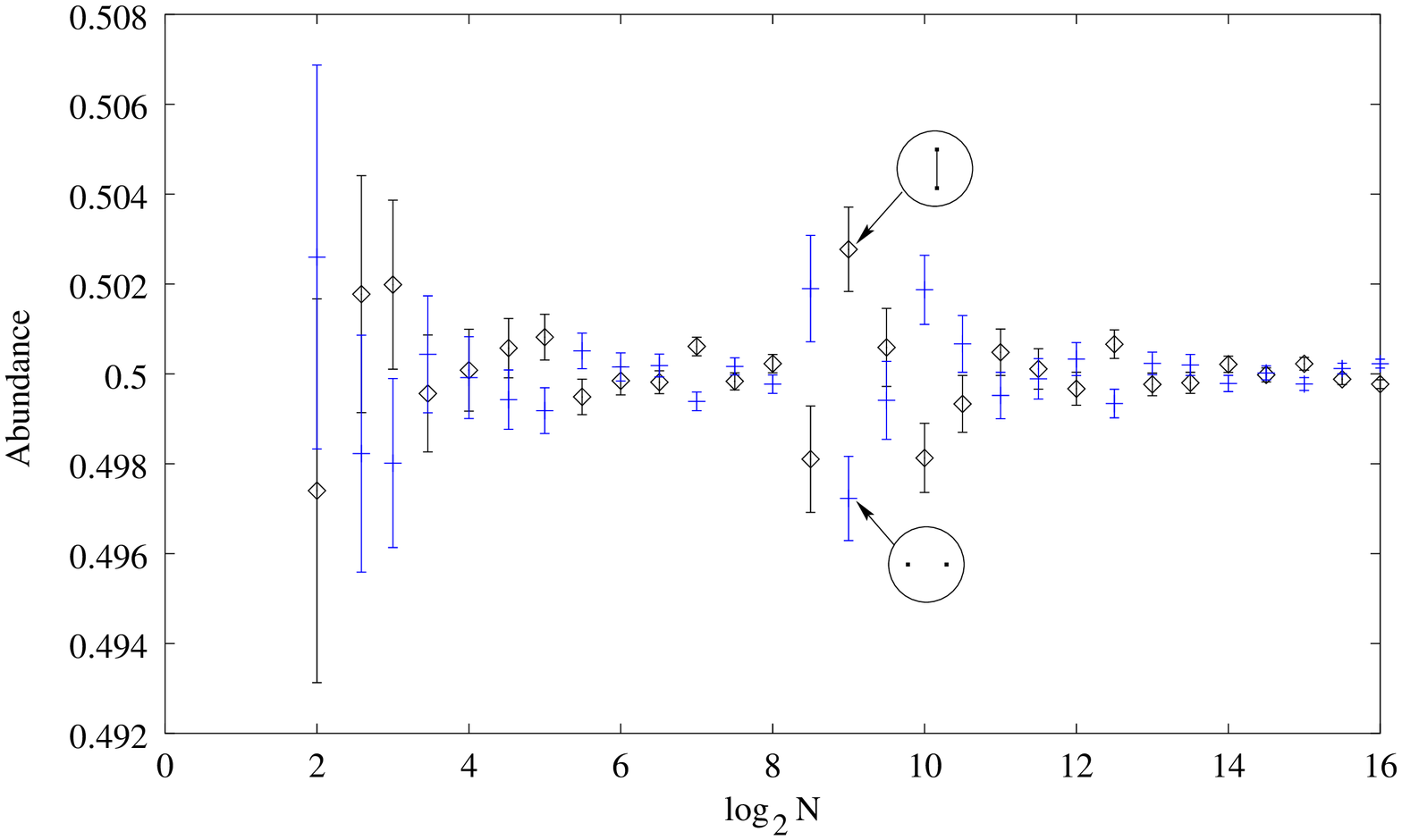} }
\caption{Flow of the coarse-grained probabilities $f_m$ for $m=2$.  
         The 2-chain probability is held at 1/2.}
\label{2ch-2}
\end{figure}

\begin{figure}[htbp]
\center
\scalebox{.76}{ % .78
\includegraphics{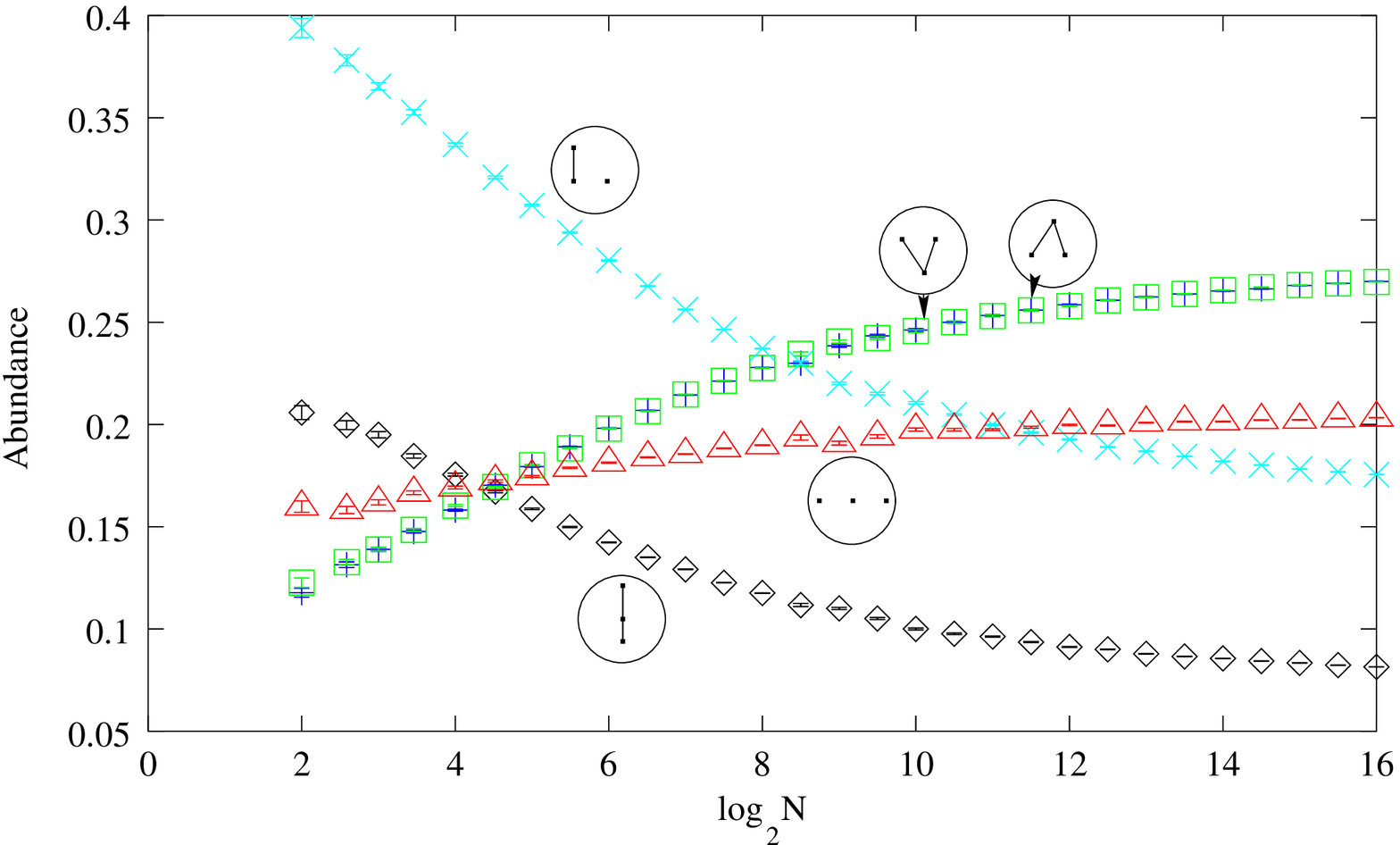} }
\caption{Flow of the coarse-grained probabilities $f_m$ for $m=3$.  
The 2-chain probability is held at 1/2.}
\label{2ch-3}
\end{figure}

\begin{figure}[htbp]
\center
\scalebox{.61}{ % .63
\includegraphics{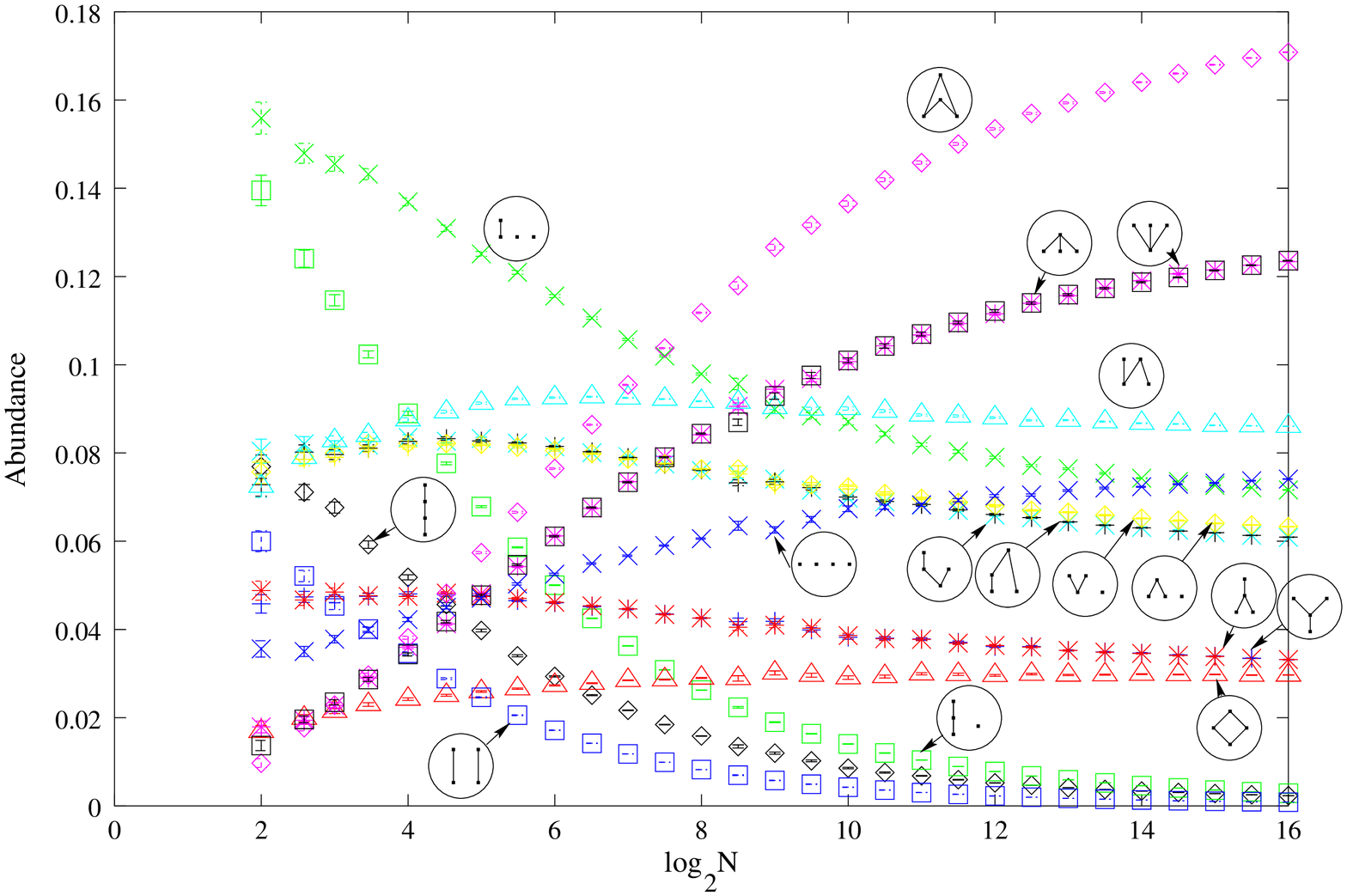} }
\caption{Flow of the coarse-grained probabilities $f_m$ for $m=4$.
         The 2-chain probability is held at 1/2.}
\label{2ch-4}
\end{figure}

In Figures \ref{2ch-2}, \ref{2ch-3}, and \ref{2ch-4}, we exhibit how
the coarse-grained probabilities of all possible 2, 3, and 4 element
causets vary as we follow the trajectory along which the
coarse-grained 2-chain probability $f_2(\twoch)=r$ is held at $1/2$.
By design, the coarse-grained probability for the 2-chain remains flat
at 50\%, so Figure \ref{2ch-2} simply shows the accuracy with which
this was achieved.
(Observe the scale on the vertical axis.)
Notice that, since 
$f_2(\twoch)$ and $f_2(\twoach)$
must sum to 1, their error bars are necessarily equal.
(The standard deviation in the abundances decreases
with increasing $N$.  The ``blip'' around $\log_2N=9$ occurs simply
because we generated fewer causets at that and larger values of $N$ to
reduce computational costs.)  

The crucial question is whether the
probabilities for the three and four element causets tend to definite
limits as $N$ tends to infinity.
Several features of the diagrams indicate that this is indeed
occurring.  Most obviously, all the curves, except possibly a couple
in Figure \ref{2ch-4}, appear to be leveling off at large $N$.  But we
can bolster this conclusion by observing in which direction the curves
are moving, and considering their interrelationships.

For the moment let us focus our attention on figure \ref{2ch-3}.
A priori there are five coarse-grained probabilities to be followed.  
That they must add up to unity reduces the degrees of freedom to four.
This is reduced further to three by the observation that, due to the
time-reversal symmetry of the percolation dynamics, we must have
$f_3(\V)=f_3(\wedge)$, as duly manifested in their graphs.  
Moreover,
all five of the curves appear to be monotonic, with 
the curves for $\wedge$, $\V$ and $\threeach$ rising, 
and
the curves for $\threech$ and $\Lcauset$ falling.
If we accept this indication of monotonicity from the diagram, then 
first of all,
every probability $f_3(\xi)$ must converge to some limiting value,
because monotonic bounded functions always do; and
some of these limits must be nonzero, because
the probabilities must add up to 1.   
Indeed, since $f_3(\V)$ and $f_3(\wedge)$ are rising, 
they must converge to some nonzero value, 
and this value must lie below 1/2 in order that the total
probability not exceed unity.  
In consequence,
the rising curve $f_3(\threeach)$ must also converge to a
nontrivial probability (one which is neither 0 nor 1).
Taken all in all, 
then, 
it looks very much like 
the $m=3$ coarse-grained theory has a nontrivial $N\to\infty$ limit, 
with at least three out of its five probabilities 
converging to nontrivial values.

Although the ``rearrangement'' of the coarse-grained probabilities
appears much more dramatic in Figure \ref{2ch-4}, 
similar arguments can be made.  
Excepting initial ``transients'', 
it seems reasonable to conclude from the data that monotonicity 
will be maintained.   
{}From this, it would follow that
the probabilities for $\flower$ and $\iflower$ 
(which must be equal by time-reversal symmetry)
and the other rising probabilities, 
$\bowtie$, $\fourach$, and $\diamond$, 
all approach nontrivial limits.  
The coarse-graining
to 4 elements, therefore,
would also admit a continuum limit 
with a minimum of 4 out of the 11 independent probabilities being nontrivial.

To the extent that the $m=2$ and $m=3$ cases are indicative, then, it is
reasonable to conclude that percolation dynamics admits a continuum limit
which is non-trivial at all ``scales'' $m$.

\begin{figure}[htbp]
\center
\scalebox{.67}{ % .69
\includegraphics{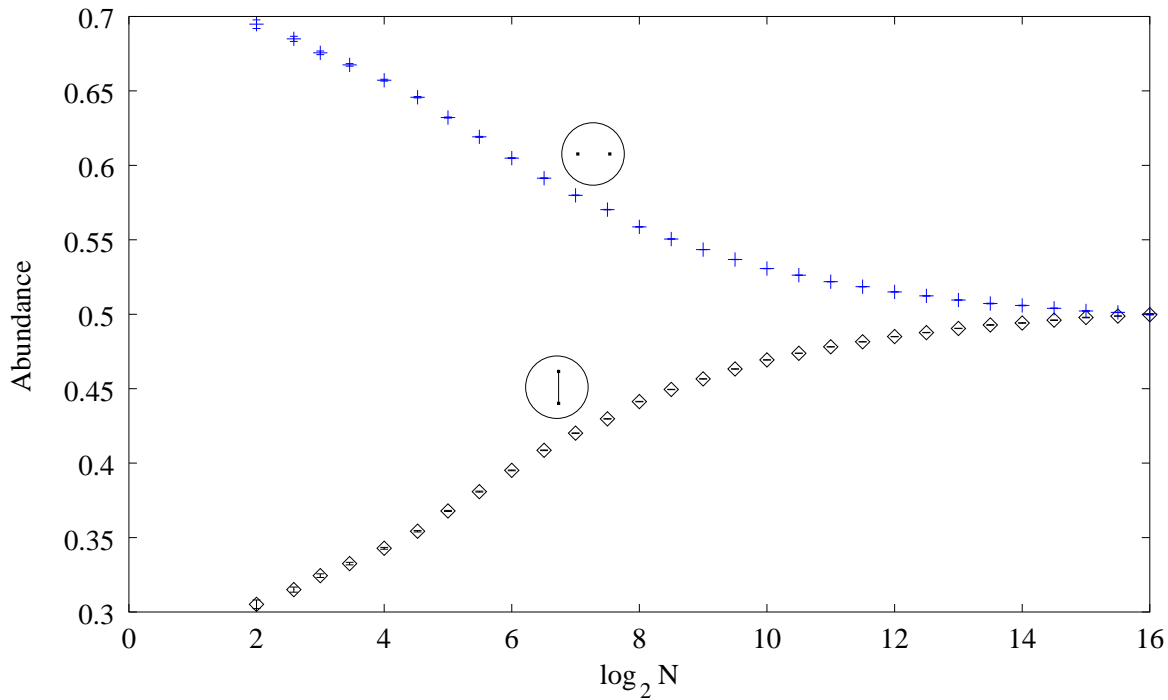} }
\caption{Flow of the coarse-grained probabilities $f_m$ for $m=2$.
         The 3-chain probability is held at 0.0814837.}
\label{3ch-2}
\end{figure}

\begin{figure}[htbp]
\center
\scalebox{.67}{ % .7
\includegraphics{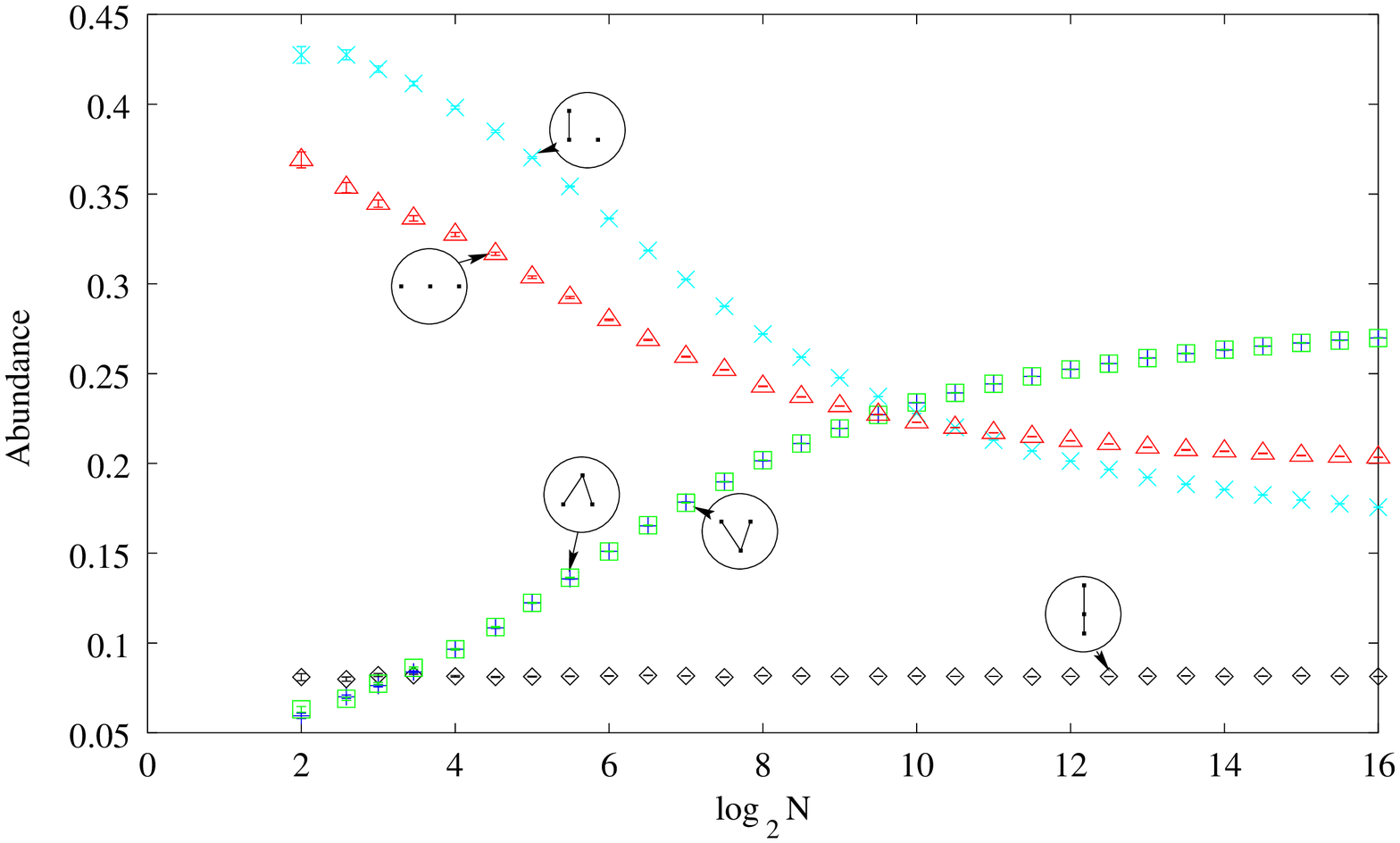} }
\caption{Flow of the coarse-grained probabilities $f_m$ for $m=3$.
        The 3-chain probability is held at 0.0814837.}
\label{3ch-3}
\end{figure}

\begin{figure}[htbp]
\center
\scalebox{.64}{ % .68
\includegraphics{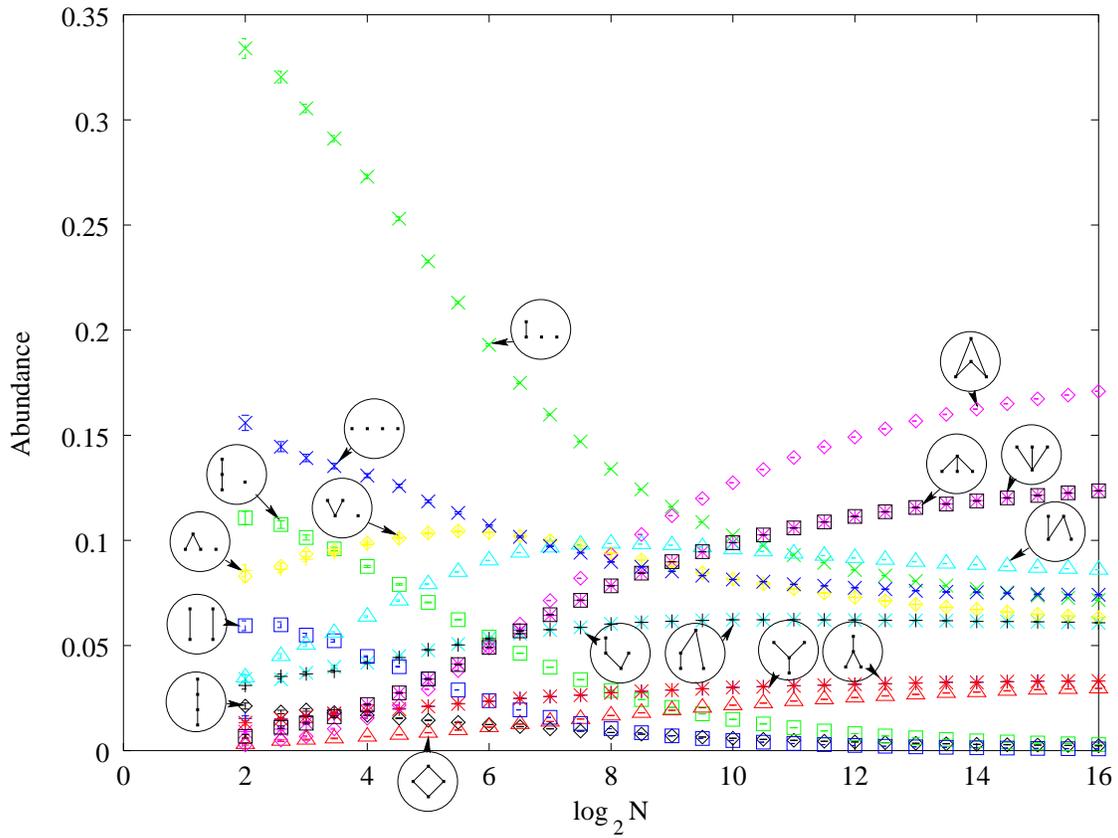} }
\caption{Flow of the coarse-grained probabilities $f_m$ for $m=4$.
         The 3-chain probability is held at 0.0814837.}
\label{3ch-4}
\end{figure}

The question suggests itself,  
whether the flow of the coarse-grained probabilities 
would differ qualitatively 
if we held fixed some abundance other than that of the 2-chain.  
In Figures \ref{3ch-2}, \ref{3ch-3}, and \ref{3ch-4},
we display results obtained by fixing the 3-chain abundance
(its value having been chosen to make the abundance of 2-chains be 1/2 
when $N=2^{16}$). 
Notice in Figure \ref{3ch-2} that 
the abundance of 2-chains varies considerably along this trajectory, 
whilst that of the 3-chain (in figure \ref{3ch-3}) 
of course remains constant.  Once again, 
the figures suggest strongly that
the trajectory is approaching a continuum limit 
with nontrivial values for 
the coarse-grained probabilities of 
at least
the 3-chain, the ``V'' and the ``$\Lambda$'' 
(and in consequence of the 2-chain and 2-antichain).

\begin{figure}[htbp]
\center
\scalebox{.63}{ % .68
\includegraphics{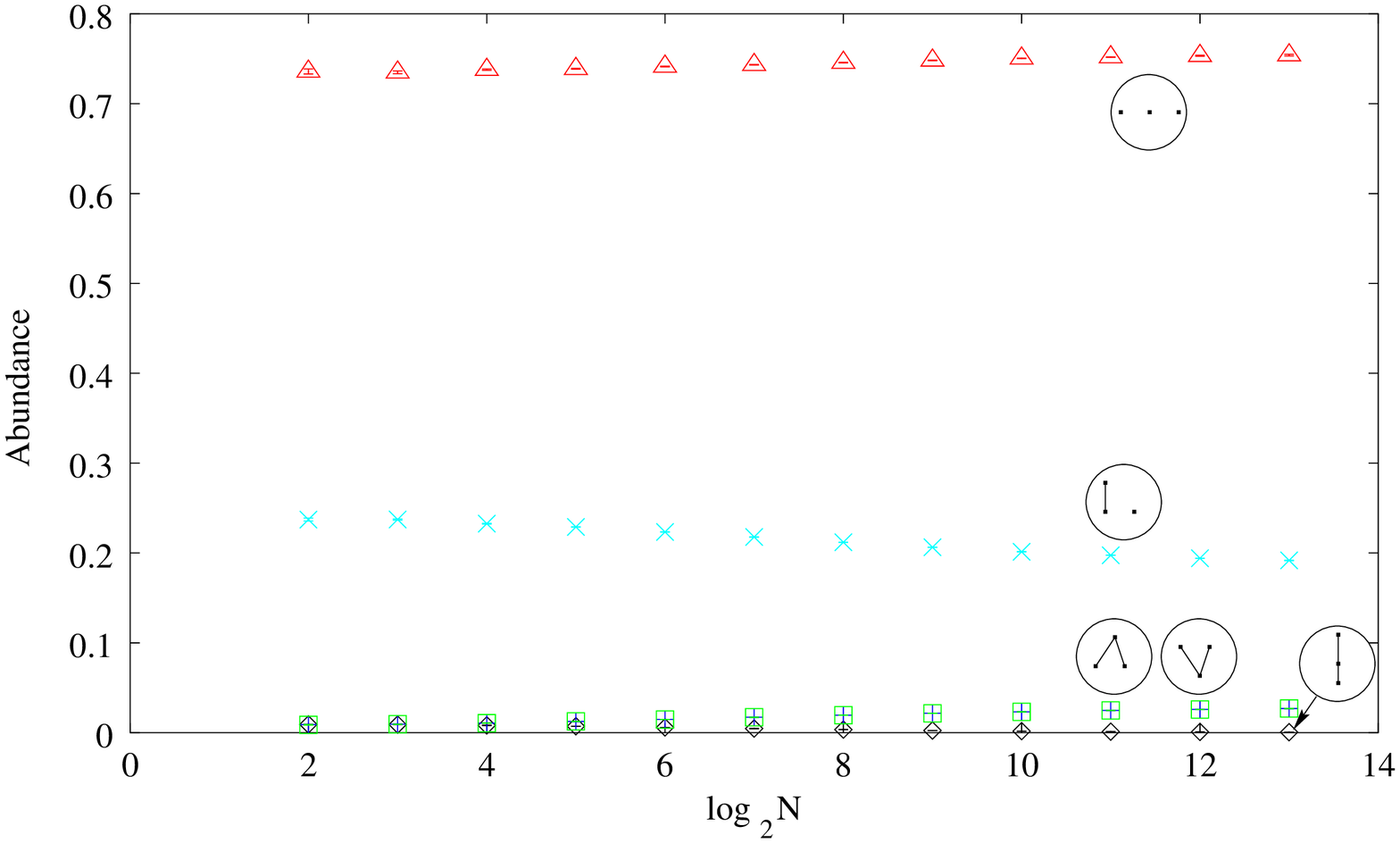} }
\caption{Flow of the coarse-grained probabilities $f_m$ for $m=3$.
         The 2-chain probability is held at 1/10.}
\label{4d-3}
\end{figure}

\begin{figure}[htbp]
\center
\scalebox{.68}{ % .68
\includegraphics{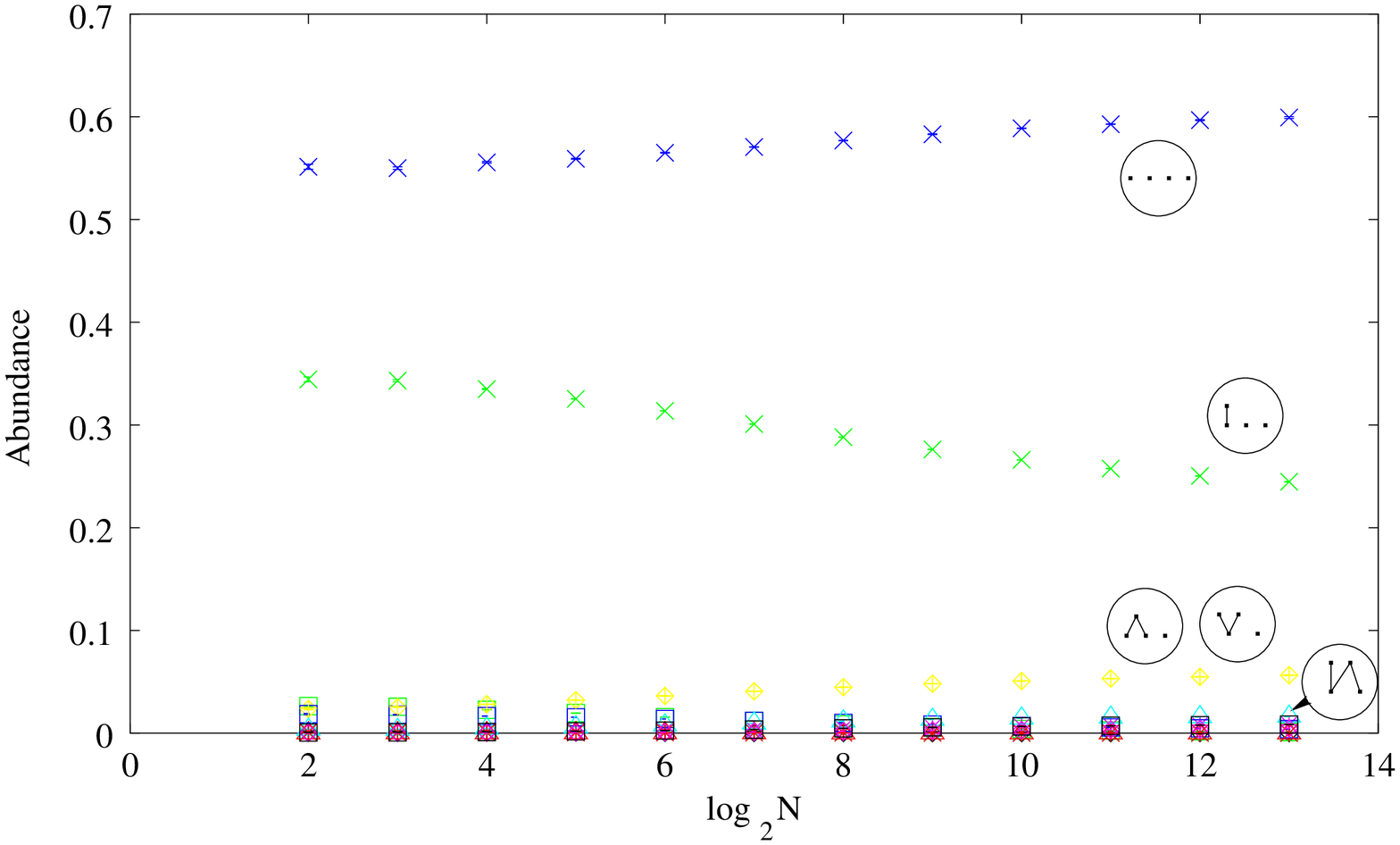} }
\caption{Flow of the coarse-grained probabilities $f_m$ for $m=4$.
      The 2-chain probability is held at 1/10.
      Only those curves lying high enough to be seen distinctly have
      been labeled.}
\label{4d-4}
\end{figure}

All the trajectories discussed so far produce causets with an ordering
fraction $r$ close to 1/2 for large $N$.
As mentioned earlier, $r=1/2$ corresponds to
a Myrheim-Meyer dimension of two.  
Figures \ref{4d-3} and \ref{4d-4} 
show the results of a simulation along the ``four dimensional'' trajectory
defined by $r=1/10$.  
(The value $r=1/10$ corresponds to a Myrheim-Meyer dimension of 4.)
Here the appearance of the flow is much less elaborate,
with the curves arrayed simply in order of increasing ordering fraction, 
$\threeach$ and $\fourach$ being at the top
and $\threech$ and (imperceptibly) $\fourch$ at the bottom.  
As before, all the curves are monotone as far as can be seen.
Aside from the intrinsic interest of the case $d=4$, these results 
indicate that our conclusions drawn for $d$ near 2 will 
hold good
for all larger $d$ as well.

\begin{figure}[htbp]
\center
\scalebox{.53}{ % .52
\includegraphics{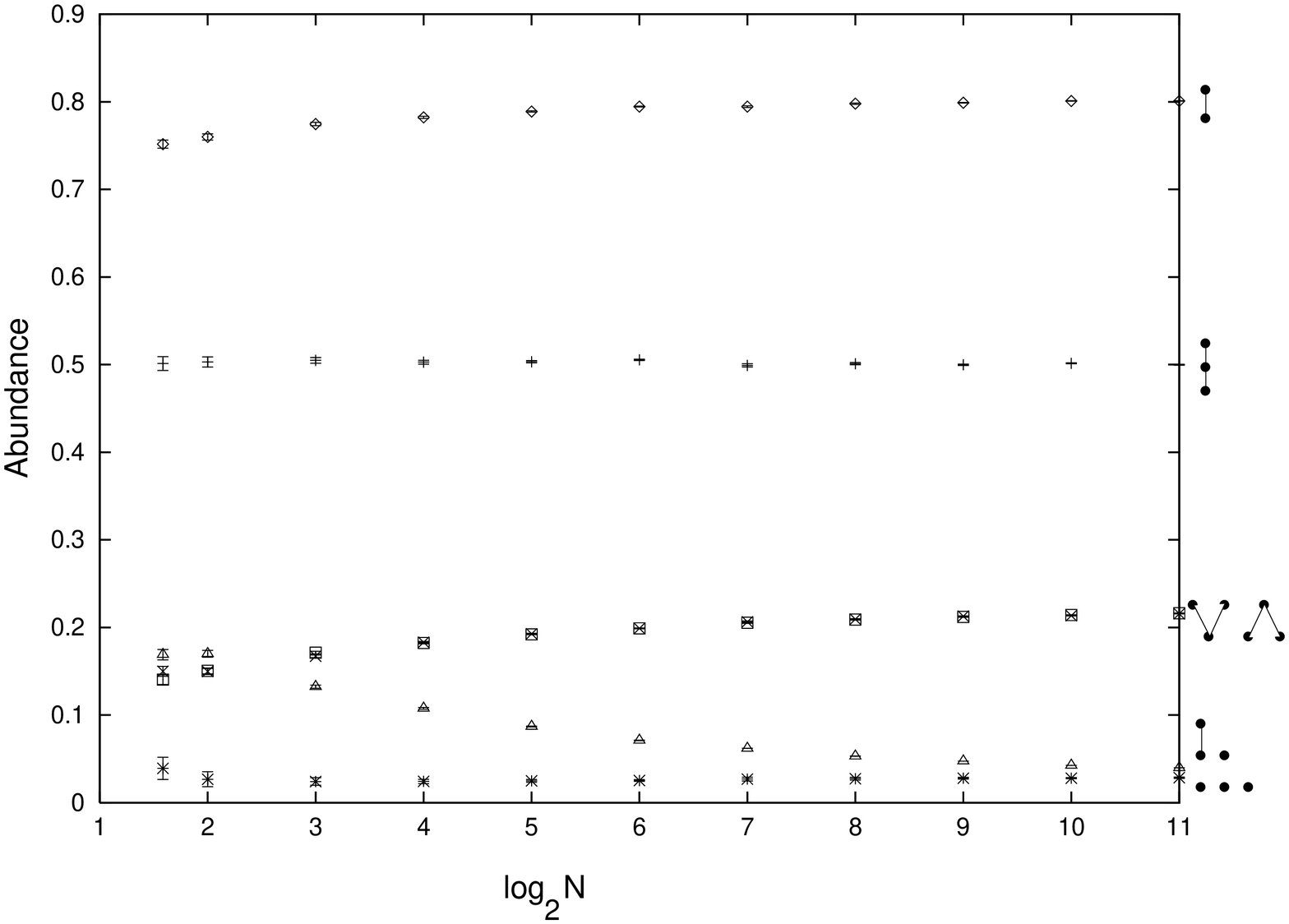} }
\caption{Flow of the coarse-grained probabilities $f_m$ for $m=3$.
         The 3-chain probability is held at 1/2.}
\label{3sub}
\end{figure}

Figure \ref{3sub} displays the flow of the coarse-grained probabilities 
from a simulation in the opposite situation where the ordering fraction
is much greater than 1/2 (the Myrheim-Meyer dimension is down near 1.)
Shown are the results of
coarse-graining to three element causets along the trajectory which
holds the 3-chain probability to 1/2.  Also shown is the 2-chain
probability.  
The behavior is similar to that of Figure \ref{4d-3},
except that here the coarse-grained probability
rises with the ordering fraction instead of falling.  
This occurs because constraining 
$f_3(\threech)$ to be 1/2
generates rather chain-like causets 
whose Myrheim-Meyer dimension is 
in the neighborhood of 1.34, 
as follows from the approximate limiting value $f_2(\twoch)\approx0.8$.  
The slow, monotonic, variation of the probabilities at large $N$, 
along with the appearance of convergence to non-zero values in each
case,  suggests the presence of a nontrivial
continuum limit for $r$ near unity as well.

\begin{figure}[htbp]
\center
\scalebox{1.07}{ % 1.2
\includegraphics{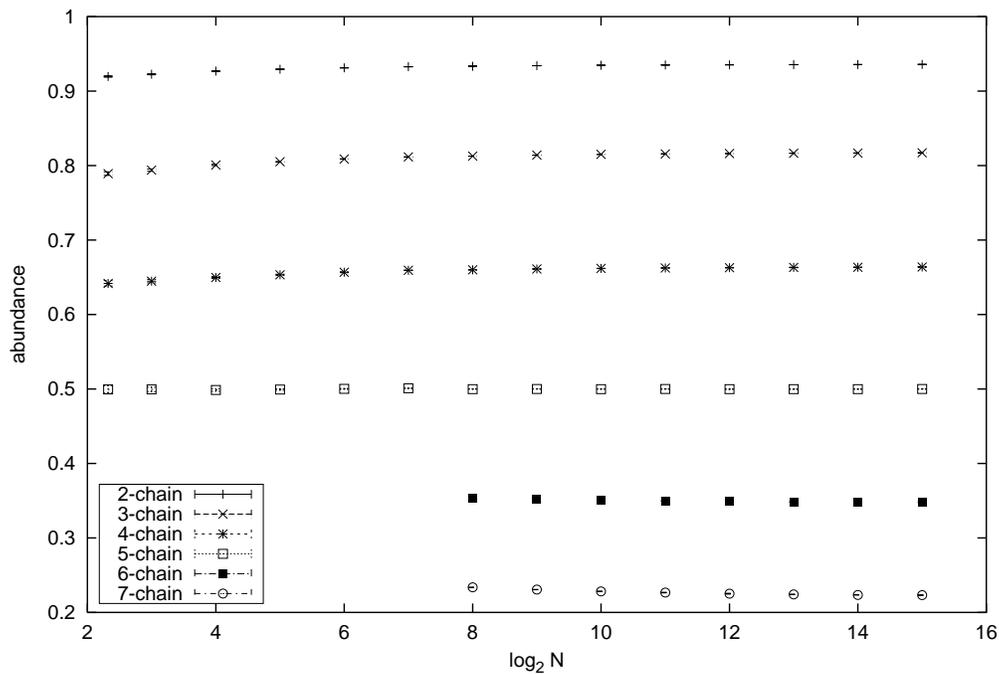} }
\caption{Flow of the coarse-grained probabilities $f_m$($m-$chain) for
   $m=2$ to $7$.  The 5-chain probability is held at 1/2.}
\label{5chtraj}
\end{figure}

\begin{figure}[htbp]
\center
\scalebox{.68}{ % .7
\includegraphics{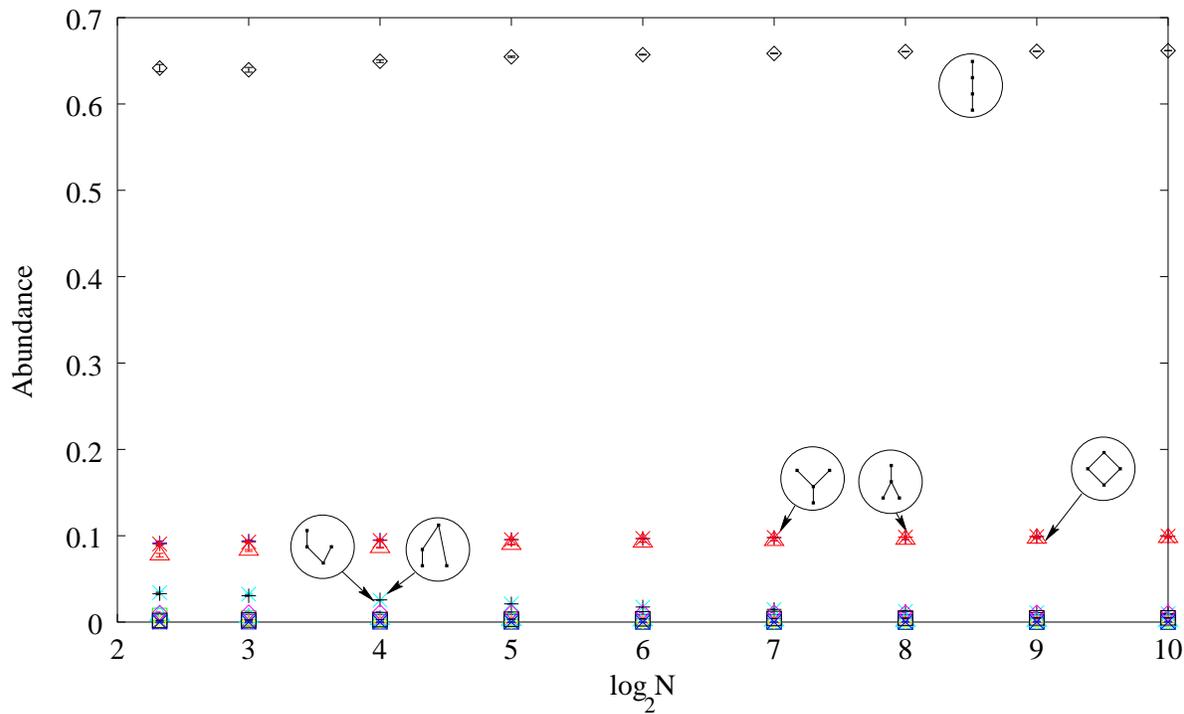} }
\caption{Flow of the coarse-grained probabilities $f_m$ for $m=4$.
         The 5-chain probability is held at 1/2.}
\label{5ch-4}
\end{figure}

Figures \ref{5chtraj} and \ref{5ch-4} present the results of a final set
of simulations, the only ones we have carried out which examined the
abundances of causets containing more than four elements.  
In these simulations, the 
mean 5-chain abundance $f_5(\hbox{\rm 5-chain})$ 
was held at 1/2, producing causets that were
even more chain-like than before
(Myrheim-Meyer dimension $\approx1.1$).
In Figure \ref{5chtraj}, we track
the resulting 
abundances of all $k$-chains for $k$ between 2 and 7, inclusive.  
(We limited ourselves to chains, because their abundances
are relatively easy to determine computationally.)
As in Figure \ref{3sub}, all the coarse-grained
probabilities appear to be tending monotonically to limits at large $N$.
%
%% [[RDS: david: is monotonic correct here? ]] 
%
In fact, 
they look amazingly constant over the whole range of $N$, 
from 5 to $2^{15}$.  
One may also observe that
the coarse-grained probability of a chain decreases markedly 
(and almost linearly over the range examined!) 
with its length, as one might expect.  
It appears
also that the $k$-chain curves for $k\not=5$ are ``expanding away''
from the 5-chain curve, but only very slightly.
Figure \ref{5ch-4} displays the flow of the probabilities for 
coarse-grainings to four elements.
%  along the $f_5(\hbox{\rm 5-chain})=1/2$ trajectory.
It is qualitatively similar to 
Figures \ref{4d-3}--\ref{3sub},
%% RDS: you may want to change the hard coding here 
with very flat probability curves,
and here with
a strong preference 
for causets having many relations 
over those having few.  
Comparing 
Figures \ref{5ch-4} and \ref{4d-4} 
with 
Figures \ref{3ch-4} and \ref{2ch-4}, 
we observe that
trajectories which generate causets that are rather chain-like or
antichain-like seem to produce distributions that converge
more rapidly than those along which the ordering fraction takes values
close to 1/2.

% The results graphed in Figure \ref{5chtraj} arose from holding the
% mean 5-chain abundance $f_5(\hbox{\rm 5-chain})$ constant at 1/2, and
% tracking the abundances of the $k$-chains for $k$ between 2 and 7,
% inclusive.  

In the way of further simulations, 
it would be extremely interesting to look for continuum limits of some
of the more general dynamical laws discussed 
in \S 4.5 of
Reference~\cite{class_dyn}.  In doing so, however, one would no longer
have available (as one does have for transitive percolation) a very
fast (yet easily coded) algorithm that generates causets randomly in
accord with the underlying dynamical law.  
Since the sequential growth
dynamics of~\cite{class_dyn} is produced by a stochastic process
defined recursively on the causal set, it is easily mimicked
algorithmically; but the most obvious algorithms that do so are too
slow to generate efficiently causets of the size we have discussed in
this paper.  Hence, one would either have to devise better
algorithms for generating causets ``one off'', or one would have to
use an entirely different method to obtain the mean abundances, like
Monte Carlo simulation of the random causet.

%: Conclusion

\section{Concluding Comments}

Transitive percolation is a discrete dynamical theory characterized by
a single parameter $p$ lying between $0$ and $1$.  Regarded as a
stochastic process, it describes the steady growth of a causal set by
the continual birth or ``accretion'' of new elements.  If we limit
ourselves to that portion of the causet comprising the elements born
between step $N_0$ and step $N_1$ of the stochastic process, we obtain
a model of random posets containing $N=N_1-N_0$ elements.  This is the
model we have studied in this paper.

Because the underlying process is homogeneous, 
this model does not depend on $N_0$ or $N_1$ separately, 
but only on their difference.  
It is therefore characterized by just two parameters $p$ and $N$.  
One should be aware that this truncation to a finite model is not
consistent with discrete general covariance, because 
it is the subset of elements 
with certain {\it labels} 
that has been selected out of the larger causet, 
rather than a subset characterized by 
any directly physical condition.  
Thus, we have introduced an ``element of
gauge'' and we hope that we are justified in having neglected it.
That is, we hope that the random causets produced by the
model we have actually studied are representative of
the type of suborder that one would obtain by percolating a much larger
(eventually infinite) causet and then using a label-invariant
criterion to select a subset of $N$ elements.

Leaving this question aside for now, let us imagine that our model
represents an interval (say) in a causet $C$ underlying some
macroscopic spacetime manifold.  With this image in mind, it is
natural to interpret a continuum limit as one in which $N\to\infty$
while the coarse-grained features of the interval in question remain
constant.  We have made this notion precise by defining coarse-graining
as random selection of a suborder whose cardinality $m$ measures the
``coarseness'' of our approximation.  A continuum limit then is
defined to be one in which $N$ tends to $\infty$ such that, for each
finite $m$, the induced probability distribution $f_m$ on the set of
$m$-element posets converges to a definite limit, 
the physical meaning being that 
the dynamics at the corresponding length-scale is well defined.
Now, how could our model {\it fail} to admit such a limit?  

In a field-theoretic setting, failure of a continuum limit to exist
typically means that the coarse-grained theory loses parameters as the
cutoff length goes to zero.
For example, $\lambda\phi^4$ scalar field theory
in 4 dimensions depends on two parameters, the mass $\mu$ and the
coupling constant $\lambda$.  
In the continuum limit, $\lambda$ is lost, 
although one can arrange for $\mu$ to survive.  
(At least this is what most workers believe occurs.)  
Strictly speaking, one should not say that a continuum limit fails to
exist altogether, but only that the limiting theory is poorer in
coupling constants than it was before the limit was taken.
Now in our case, 
we have only one parameter to start with, 
and we have seen that it does survive as $N\to\infty$ 
since we can, 
for example, 
choose freely 
the $m=2$ coarse-grained probability distribution $f_2$.
Hence, 
we need not fear such a loss of parameters in our case.

What about the opposite possibility?  
Could the coarse-grained theory {\it gain} parameters 
in the $N\to\infty$ limit, 
as might occur 
if the distributions $f_m$ were sensitive to 
the fine details of 
the trajectory along which $N$ and $p$ 
approached the ``critical point'' $p=0$, $N=\infty$?\footnote%
{Such an increase of the parameter set through a limiting process
 seems logically possible,
 although we know of no example of it from field theory or statistical
 mechanics, unless one counts the extra global parameters that come in with 
 ``spontaneous symmetry breaking''.}
%
%  For example, a water-ice equilibrium is not characterized by the 
%  temperature and pressure alone, though it is by density and temperature.
%
Our simulations showed no sign of such sensitivity, 
although we did not look for it specifically.
(Compare, for example, Figure 10 with Figure 13 and 11 with 14.)
%
% Rather, trajectories defined with respect
% to different reference causets behave qualitatively very much alike,
% as one sees by comparing Figure \ref{3sub}, which used the 3-chain as
% reference poset, with Figure \ref{5chtraj}, which used the 5-chain.

A third way the continuum limit could fail might perhaps be viewed as an
extreme form of the second.  It might happen that, no matter how one chose
the trajectory $p=p(N)$, 
some of the coarse-grained probabilities $f_m(\xi)$ 
oscillated indefinitely as $N\to\infty$, 
without ever settling down to fixed values.  
Our simulations leave little
room for this kind of breakdown, since they manifest the exact opposite
kind of behavior,
namely monotone variation of all the 
coarse-grained probabilities we ``measured''.

Finally, 
a continuum limit could exist in the technical sense, 
but it still could be effectively trivial 
(once again reminiscent of the $\lambda\phi^4$ case --- 
if you care to regard a free field theory as trivial.)  
Here triviality would mean that all --- or almost all --- of 
the coarse-grained probabilities $f_m(\xi)$ 
converged either to 0 or to 1.  
Plainly, we can avoid this for at least some of the $f_m(\xi)$.  
For example, we could choose an $m$ and hold either
$f_m(m\hbox{-chain})$ or $f_m(m\hbox{-antichain})$ fixed
at any desired value.
(Proof: as $p\to1$, $f_m(m\hbox{-chain})\to1$ and
$f_m(m\hbox{-antichain})\to0$; as $p\to0$, the opposite occurs.)
However, in principle, it could still happen that all the other $f_m$
besides these two went to 0 in the limit.  
(Clearly, they could not go to 1, the other trivial value.)  
Once again, our simulations show the opposite behavior.
For example, we saw that $f_3(\V)$ 
{\it increased} monotonically
along the trajectory of Figure \ref{2ch-3}.

Moreover, even without reference to the simulations, we can make this
hypothetical ``chain-antichain degeneracy'' appear very implausible by
considering a ``typical'' causet $C$ generated by percolation for
$N\gg 1$
with $p$ on the trajectory that, for some chosen $m$, holds
$f_m(m\hbox{-chain})$ fixed at a value $a$ strictly between 0 and 1.
Then our degeneracy would insist that $f_m(m\hbox{-antichain})=1-a$ and
$f_m(\chi)=0$ for all other $\chi$.  But this would mean that, in a
manner of speaking, ``every'' coarse-graining of $C$ to $m$ elements
would be either a chain or an antichain.  In particular the causet
\Lcauset could not occur as a subcauset of $C$; whence, since \Lcauset
is a subcauset of every $m$-element causet except the chain and the
antichain, $C$ itself would have to be either an antichain or a chain.
But it is absurd that percolation for any parameter value $p$ other than
0 and 1 would produce a ``bimodal'' distribution such that $C$ would
have to be either a chain or an antichain, but nothing in between.  (It
seems likely that similar arguments could be devised against the
possibility of similar, but slightly less trivial trivial continuum
limits, for example a limit in which $f_m(\chi)$ would vanish unless
$\chi$ were a disjoint union of chains and antichains.)

Putting all this together, we have persuasive evidence that the
percolation model does admit a continuum limit, 
with the limiting model being 
nontrivial and 
described by a single ``renormalized'' parameter or ``coupling constant''.  
Furthermore, the associated scaling behavior
one might anticipate in such a case is also present, as we will
discuss further in \cite{scaling}.  
% 
% In this sense, the percolation model may be said to be ``renormalizable''.  
% Oops: ``renormalizable'' used non-standardly here.  sentence deleted. 

But is the word ``continuum'' here just a metaphor, or can it be taken
more literally?  This depends, of course, on the extent to which the
causets yielded by percolation dynamics resemble genuine spacetimes.
Based on the meager evidence available at the present time, we can only
answer ``it is possible''.  On one hand, we know \cite{class_dyn} that
any spacetime produced by percolation would have to be homogeneous, like
de Sitter space or Minkowski space. We also know, from simulations in
progress, that two very different dimension estimators seem to agree on
percolated causets, which one might not expect, were there no actual
dimensions for them to be estimating.  Certain other indicators tend to
behave poorly, on the other hand, but they are just the ones that are
{\it not} invariant under coarse-graining (they are not ``RG
invariants''), so their poor behavior is consistent with the expectation
that the causal set will not be manifold-like at the smallest scales
(``foam''), but only after some degree of coarse-graining.  

Finally, there is the ubiquitous issue of ``fine tuning'' or ``large
numbers''.  In any continuum situation, a large number is 
being
manifested
(an actual infinity in the case of a true continuum) and one may
wonder where it came from.  In our case, the large numbers were
$p^{-1}$ and $N$.  For $N$, there is no mystery: unless the birth
process ceases, $N$ is guaranteed to grow as large as desired.  But why
should $p$ be so small?  Here, perhaps, we can appeal to the
preliminary results of Dou mentioned in the introduction.  If ---
cosmologically considered --- the causet that is our universe has
cycled through one or more phases of expansion and recollapse, then
its dynamics will have been filtered through a kind of ``temporal
coarse-graining'' or ``RG transformation'' that tends to drive it
toward transitive percolation.  But what we didn't mention earlier was
that the parameter $p$ of this effective dynamics scales like
$N_0^{-1/2}$, where $N_0$ is the number of elements of the
causet preceding the most recent ``bounce''.  Since this is sure to
be an enormous number if one waits long enough, $p$ is sure to become
arbitrarily small if sufficiently many cycles occur.  The reason for
the near flatness of spacetime --- or if you like for the large
diameter of the contemporary universe --- would then be just that the
underlying causal set is very old --- old enough to have accumulated,
let us say, $10^{480}$ elements in earlier cycles of expansion,
contraction and re-expansion.

% %%%%%%%%%%%%%

%: Acknowledgements

It is a pleasure to thank Alan Daughton, Chris Stephens,
Henri Waelbroeck and Denjoe {\'O}Connor
for extensive discussions on the subject of this paper.
The research reported here was supported in part by NSF grants
PHY-9600620 and INT-9908763 and by a grant from the Office of Research
and Computing of Syracuse University.

%\bibliography{../include/references}

\end{document}